\documentclass[twoside,twocolumn,10pt]{article}

\RequirePackage[ruled,noend,noline,slide]{algorithm2e}
\makeatletter
\newenvironment{Ualgorithm}[1][htpb]
{\def\@algocf@post@ruled{\kern\interspacealgoruled\hrule  height\algoheightrule\kern3pt\relax}%
	\def\@algocf@capt@ruled{under}
	\begin{algorithm}[#1]}
	{\end{algorithm}}
\makeatother

\newcommand*\patchAmsMathEnvironmentForLineno[1]{%
	\expandafter\let\csname old#1\expandafter\endcsname\csname #1\endcsname
	\expandafter\let\csname oldend#1\expandafter\endcsname\csname end#1\endcsname
	\renewenvironment{#1}%
	{\linenomath\csname old#1\endcsname}%
	{\csname oldend#1\endcsname\endlinenomath}}%
\newcommand*\patchBothAmsMathEnvironmentsForLineno[1]{%
	\patchAmsMathEnvironmentForLineno{#1}%
	\patchAmsMathEnvironmentForLineno{#1*}}%
\AtBeginDocument{%
	\patchBothAmsMathEnvironmentsForLineno{equation}%
	\patchBothAmsMathEnvironmentsForLineno{align}%
	\patchBothAmsMathEnvironmentsForLineno{flalign}%
	\patchBothAmsMathEnvironmentsForLineno{alignat}%
	\patchBothAmsMathEnvironmentsForLineno{gather}%
	\patchBothAmsMathEnvironmentsForLineno{multline}%
}

\usepackage{wscg}           

\RequirePackage{ifpdf}
\ifpdf
\RequirePackage[pdftex]{graphicx}
\else
\RequirePackage[dvips,draft]{graphicx}
\fi

\RequirePackage{lineno}

\usepackage{fancyhdr}

\usepackage{amssymb}
\usepackage{amsmath}
\usepackage [nolist]{acronym}
\usepackage{verbatim}

\def\permil{\ensuremath{{}^\text{o}\mkern-5mu/\mkern-3mu_\text{oo}}}

\usepackage{nopageno}       

\usepackage{subcaption}

\ifdefined\review
\fi

\title{Accelerating Evolutionary Construction Tree Extraction via Graph Partitioning}

\ifdefined\review
\author{
	\parbox{0.25\textwidth}{\centering
		First Author\\[1mm]
		author's affiliation\\
		1st line of address\\
		2nd line of address\\
		Country (ZIP) code, City, State\\[1mm]
		e@mail
	}
	\hspace{0.05\textwidth}
	\parbox{0.25\textwidth}{\centering
		Second Author\\[1mm]
		author's affiliation\\
		1st line of address\\
		2nd line of address\\
		Country (ZIP) code, City, State\\[1mm]
		e@mail
	}
	\hspace{0.05\textwidth}
	\parbox{0.25\textwidth}{\centering
		Third Author\\[1mm]
		author's affiliation\\
		1st line of address\\
		2nd line of address\\
		Country (ZIP) code, City, State\\[1mm]
		e@mail
	}
}
\else 
\author{
	\parbox{0.5\textwidth}{\centering
		Markus Friedrich, Sebastian Feld, Thomy Phan\\[1mm]
		Institute for Computer Science\\
		Ludwig-Maximilians-University Munich\\
		Oettingenstr. 67\\
		80538 Munich, Germany\\[1mm]
		\{markus.friedrich|sebastian.feld|thomy.phan\}@ifi.lmu.de
	}
	\hspace{0.0\textwidth}
	\parbox{0.50\textwidth}{\centering
		Pierre-Alain Fayolle\\[1mm]
		Division of Information and Systems\\
		The University of Aizu\\
		Aizu-Wakamatsu City \\
		965-8580 Fukushima, Japan\\[1mm]
		fayolle@u-aizu.ac.jp
	}
}
\fi

\usepackage{url}
\makeatletter
\def\Uslash{\mathbin{\mathchar`\/}\@ifnextchar{/}{\kern-.15em}{}}
\g@addto@macro\UrlSpecials{\do \/ {\Uslash}}
\def\Ucolon{\mathbin{\mathchar`:}\@ifnextchar{/}{\kern-.1em}{}}
\g@addto@macro\UrlSpecials{\do : {\Ucolon}}
\makeatother

\begin{document}
	
	\ifdefined\review
		\setpagewiselinenumbers	
		\modulolinenumbers[5]
		\linenumbers
		\setlength\linenumbersep{5pt}	
	\fi

	\twocolumn[{\csname @twocolumnfalse\endcsname
		
		\maketitle  
		
		\begin{abstract}
			\noindent
			Extracting a Construction Tree from potentially noisy point clouds is an important aspect of Reverse Engineering tasks in Computer Aided Design. 
Solutions based on algorithmic geometry impose constraints on usable model representations (e.g. quadric surfaces only) and noise robustness. 
Re-formulating the problem as a combinatorial optimization problem and solving it with an Evolutionary Algorithm can mitigate some of these constraints at the cost of increased computational complexity. 
This paper proposes a graph-based search space partitioning scheme that is able to accelerate Evolutionary Construction Tree extraction while exploiting parallelization capabilities of modern CPUs.
The evaluation indicates a speed-up up to a factor of $46.6$ compared to the baseline approach while resulting tree sizes increased by $25.2\%$ to $88.6\%$. 

		\end{abstract}
		\subsection*{Keywords}
		$3$-d Reconstruction, Reverse Engineering, Computer Aided Design, Constructive Solid Geometry, Evolutionary Algorithms, Graph Theory
		\vspace*{1.0\baselineskip}
	}]

	\begin{acronym}
		\acro{CSG}{Constructive Solid Geometry}
		\acro{CAD}{Computer Aided Design}
		\acro{GA}{Genetic Algorithm}
		\acro{RE}{Reverse Engineering}
		\acro{BRep}{Boundary Representation}
		\acro{PO}{Primitive Overlap}	
		\acro{AABB}{Axis-Aligned Bounding Box}
		\acrodefplural{AABB}[AABBs]{Axis-Aligned Bounding Boxes}
		\acro{OBB}{Oriented Bounding Box}
		\acrodefplural{OBB}[OBBs]{Oriented Bounding Boxes}	
		\acro{BKA}{Bron-Kerbosch Algorithm}
		\acro{RANSAC}{Random Sample Consensus}
	\end{acronym}

\section{Introduction}
\ac{RE} -- i.e., the recovery of a model's geometric representation from potentially noisy and incomplete sensor data -- is an important aspect of modern \ac{CAD} pipelines. 
It allows for convenient model editing based on real-world physical objects, thus simplifying and accelerating the product design process.
An expressive and intuitive model representation scheme extensively used in solid modeling is \ac{CSG}.
It describes complex rigid solids by a binary tree with regularized Boolean set-operations (e.g., union, intersection, subtraction) as inner nodes and primitive solids (e.g., cubes, spheres, cylinders and cones) as leaves. 
\copyrightspace
Such a tree is also known as a model's Construction Tree.
Due to the popularity of \ac{CSG} in \ac{CAD}, it is desirable to have tools at hand that are able to reliably recover a model's \ac{CSG}-tree from its point cloud representation stemming from sensor recordings.
\ac{CSG}-tree generation might be solved by converting the input point cloud to a Boundary Representation (B-rep) and then by conversion of the B-rep to CSG with methods based on algorithmic geometry that usually require exact geometric intersection computations \cite{shapiro1993separation, buchele2004three}. 
These approaches are usually restricted to a single model representation for primitives, e.g. a surface description that uses quadrics, and can be sensitive to inexact representations. 
\\
To overcome these constraints, \ac{CSG}-tree generation can be formulated as a combinatorial optimization problem over the possible permutations of primitives and set-operations for a fixed maximum \ac{CSG}-tree depth.
Metaheuristics, like \acp{GA} can then be employed for optimization \cite{mitchell1998introduction}.
\\
One of the most severe disadvantages of \ac{GA}-based solutions are computation times of minutes and hours for comparably small models (less than $10$ primitives) \cite{fayolle2016evolutionary}.
This issue is addressed in this paper.  
\\
The basic idea of the described acceleration scheme is to exploit spatial relationships between primitives: 
Primitives that do not overlap spatially are not considered to be operands of a \ac{CSG}-operation.
This knowledge can be used to partition overlapping primitives and to compute partial per-partition results that are later on merged into a single \ac{CSG}-tree. 
In particular, this paper makes the following contributions:
\begin{itemize}
\item An acceleration scheme based on spatial search space partitioning together with a robust merge mechanism.
\item A description and analysis of parallelization strategies for the proposed algorithms.
\end{itemize}  
The paper has the following structure: 
Section \ref{sec:relworks} discusses related work in the field of \ac{CSG}-tree extraction and surface reconstruction.
It is followed by an introduction to the theoretical principles of the proposed method (Section \ref{sec:back}).
The problem to solve is detailed in Section \ref{sec:prob}.
The proposed solution is described in Section \ref{sec:concept} and evaluated in Section \ref{sec:eval}.
Section \ref{sec:conclusion} summarizes the results and sketches possible future work.

\section{Related Work}
\label{sec:relworks}
This work is related to different domains such as surface reconstruction from discrete point clouds, Reverse Engineering of solid models and conversion from B-rep to \ac{CSG}. 
In this section, important related work in these domains is briefly discussed. 

\subsection{Surface Reconstruction}
The problem of reconstructing a surface from a discrete point cloud has been the subject of much attention in computer graphics. 
The most popular methods include fitting implicit surfaces such as \cite{OBATS03}, or Poisson surface reconstruction \cite{KH13}, among others. 
The recent work of Berger et al. \cite{berger2017survey} presents a wide survey of the topic.
Using these methods, the reconstructed objects lack information that can be used for inspection or re-use of the object in further modeling. 

\subsection{Reverse Engineering and B-rep to CSG Conversion}
The goal of Reverse Engineering is the creation of consistent geometric models from point cloud data \cite{VMC97,BMV01}. 
They usually output B-rep models made of parametric patches.
\\
The conversion from B-rep to CSG was first investigated 
for two-dimensional, linear polygons, then 
later extended by Shapiro et al. for handling curved polygons \cite{shapiro1991efficient, shapiro2001convex}. 
The extension to three-dimensional objects was initially solved 
by Shapiro and Vossler in 
\cite{shapiro1991construction, shapiro1993separation} 
and later improved by 
Buchele and Crawford in \cite{buchele2004three}. 
These works rely on the fact that surfaces are composed of quadric surface patches. 
Another issue is the handling of inexact representations. 
These methods work under the assumption that the patches form a clean partition of the 
target solid. However, in practice, we are dealing with input point clouds that are potentially 
noisy, contain holes, or have additional details and thus the fitted primitives may not fit perfectly. 
This could impact the cellular classification on which these methods rely. 

\subsection{Point Cloud to CSG Construction}
In \cite{xiao2014}, a greedy approach is used to build a \ac{CSG}-representation with cuboids as primitives. This approach is limited to the reconstruction of buildings. 
Close to the proposed approach are methods that handle noisy and incomplete point clouds 
such as \cite{schnabel2007efficient} for fitting primitives and methods that try to convert them to a higher level representation such as \cite{fayolle2016evolutionary}, see also \cite[Sections~7 and 8]{berger2017survey}. 
One of the goals of this work is to improve the running time of the Evolutionary Algorithm used in \cite{fayolle2016evolutionary} via geometric consideration, i.e. the overlapping in space of primitives.

\section{Background}
\label{sec:back}
\subsection{Point Cloud to \ac{CSG}-Tree Pipeline} 
\label{sec:pipeline}
The extraction of a \ac{CSG}-tree from a point cloud poses a complex problem which is usually solved with a processing pipeline that comprises the following steps:  
\begin{enumerate}
\item \textbf{Point cloud generation and pre-processing:} Point clouds are generated by laser scanners or tactile measurement devices. 
Other techniques use photogrammetric algorithms to gather depth information from (un-)calibrated camera images \cite{hartley2003multiple}.
Measured point clouds usually contain significant amounts of noise and outliers. 
These can be trimmed from the data-set using e.g. statistical approaches  \cite{rusu20113d}.
\item \textbf{Point cloud segmentation and primitive fitting:} The point cloud must be segmented and primitive parameters have to be fitted to the corresponding points. Approaches that fulfill both tasks for simple geometric shapes are e.g. specialized variants of the \ac{RANSAC} technique \cite{schnabel2007efficient}.
\item \textbf{\ac{CSG}-tree generation:} \ac{CSG}-tree generation can be done with methods based on algorithmic geometry such as \cite{shapiro1993separation, buchele2004three}, 
or via evolutionary approaches such as \cite{fayolle2016evolutionary} for handling inexact representations.
\item \textbf {\ac{CSG}-tree optimization:} The resulting \ac{CSG}-tree might not be optimal in terms of size and depth.
Additional optimization techniques can simplify the tree structure \cite{shapiro1991construction}. 
\end{enumerate}
\subsection{Primitive Description}
Primitives are basic shapes located at \ac{CSG}-tree leaves. 
A primitive $p$ is fully described by its signed distance function $f_p: \mathbb{R}^3 \mapsto \mathbb{R}$.
The surface of $p$ is implicitly defined by the zero-set of $f_p$: $\{x \in \mathbb{R}^3 : f_p(x)=0\}$.
Its surface normal at point $x \in \mathbb{R}^3$ is given by the gradient $\nabla f_p(x)$.
If the gradient does not exist at $x$ or is too expensive to compute, 
finite difference approximations can be used.
\subsection{Boolean Set-Operations}
The set-operations intersection, union, complement and subtraction are implemented using $\min$- and $\max$-functions \cite{ricci197constgeo}: 
\begin{itemize}
	\item Intersection: $S_1 \cap S_2 := \min(f_{S_1}, f_{S_2})$
	\item Union: $S_1 \cup S_2 := \max(f_{S_1}, f_{S_2})$
	\item Complement: $\overline{S} := -f_S$ 
	\item Subtraction: $S_1 \setminus S_2 := S_1 \cap \overline{S_2}$
\end{itemize}
where $S_i$ is the solid corresponding to the set $\{x \in \mathbb{R}^3: f_{S_i} \geq 0\}$ ($i=1,2$).
In the following, the considered Boolean set-operations are intersection, union, complement and subtraction.
\subsection{Evolutionary Algorithms} 
\begin{figure}[htb]
	\centering
	\includegraphics[width=0.35\textwidth]{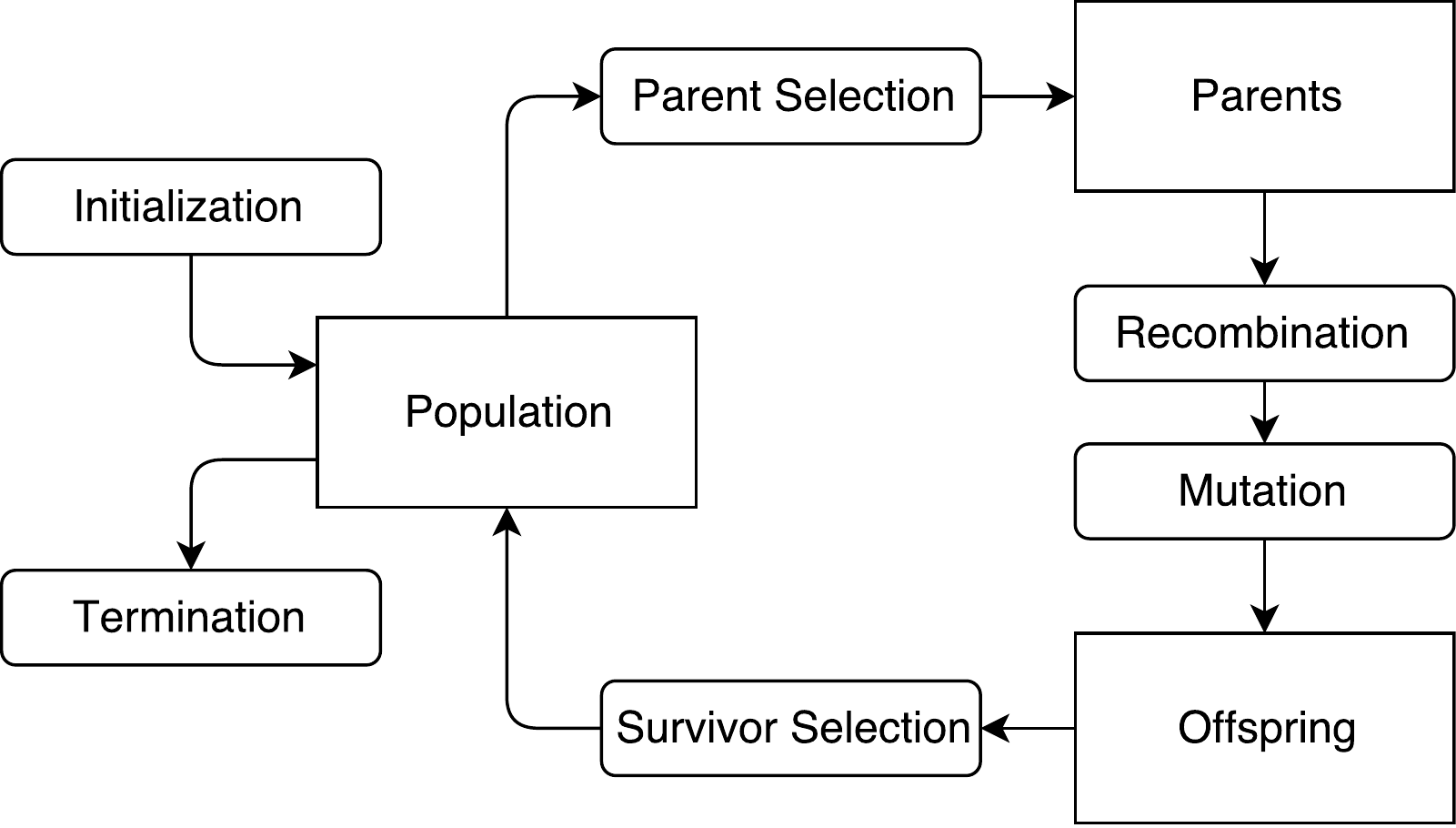}
	\caption{The optimization process described by an Evolutionary Algorithm (derived from \cite{eiben2003introduction}).}
	\label{fig:evo}
\end{figure}

Evolutionary Algorithms are biology-inspired, stochastic metaheuristics for solving optimization problems \cite{eiben2003introduction}.
\\
The optimization process starts with a randomly initialized population of individual candidates sampled from the problem's search space (initialization).
In each iteration, candidates are ranked according to their fitness by evaluating the so-called fitness function.
The best candidates are selected to be the next generation's parents (parent selection).
Parents are then recombined (crossover) and mutated (mutation) to create offspring. 
The new population is then filled with the offspring together with selected surviving individuals (survivor selection) from the current population.
This procedure is repeated until a certain termination criteria is met (termination). 
See Fig.~\ref{fig:evo} for an overview.
\\
Evolutionary Algorithms are especially useful for solving combinatorial optimization problems \cite{eiben2003introduction}.
\section{Problem Statement}
\label{sec:prob}
The problem of accelerating \ac{GA}-based \ac{CSG}-tree extraction from point clouds is considered as the open research question addressed by this paper.
The focus is on \ac{CSG}-tree generation and optimization (step $3$ and $4$ of the pipeline detailed in Section \ref{sec:pipeline}).
As input, a point-set of potentially noisy $3$-d measurements of a connected geometric model is considered. We also assume that the point-set is already segmented with fitted primitives, using techniques depicted in step $1$ and $2$ of the pipeline described in Section \ref{sec:pipeline}.
\\
The desired output is a \ac{CSG}-tree that represents the scanned real-world model as accurately as possible.
A measure for accuracy is given by the distance between the \ac{CSG}-tree induced surface and the points of the input point cloud.
\ac{CSG}-tree extraction approaches based on a \ac{GA} \cite{fayolle2016evolutionary} can handle 
inaccuracies but come with the disadvantage of potentially high computation times.
\section{Concept}
\label{sec:concept}
\begin{figure}[htb]
	\centering
	\begin{tabular}[c]{ccc}
		\begin{subfigure}[c]{0.28\linewidth}
			\includegraphics[width=\textwidth]{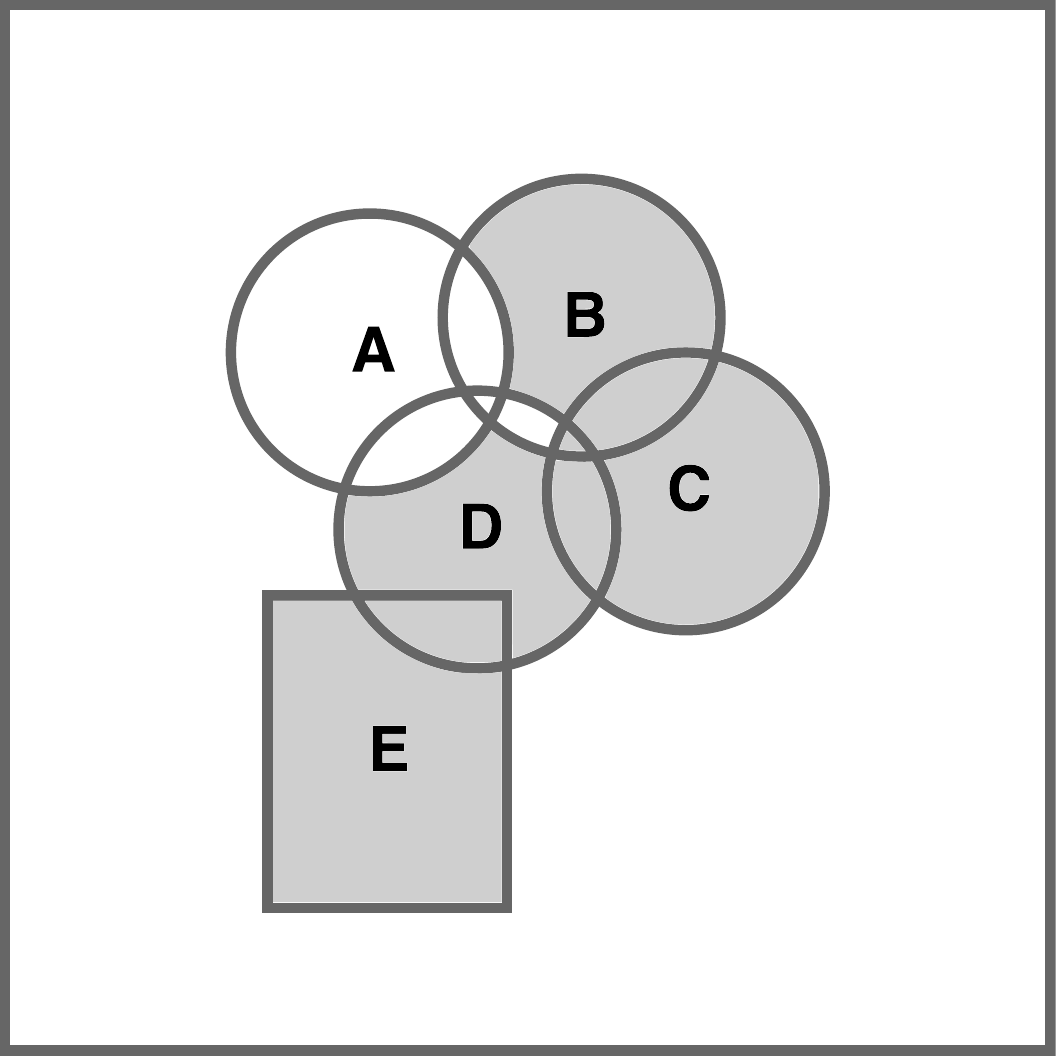}
			\caption{Primitives.}
			\label{fig:pipe0}
		\end{subfigure}&
		\begin{subfigure}[c]{0.28\linewidth}
			\includegraphics[width=\textwidth]{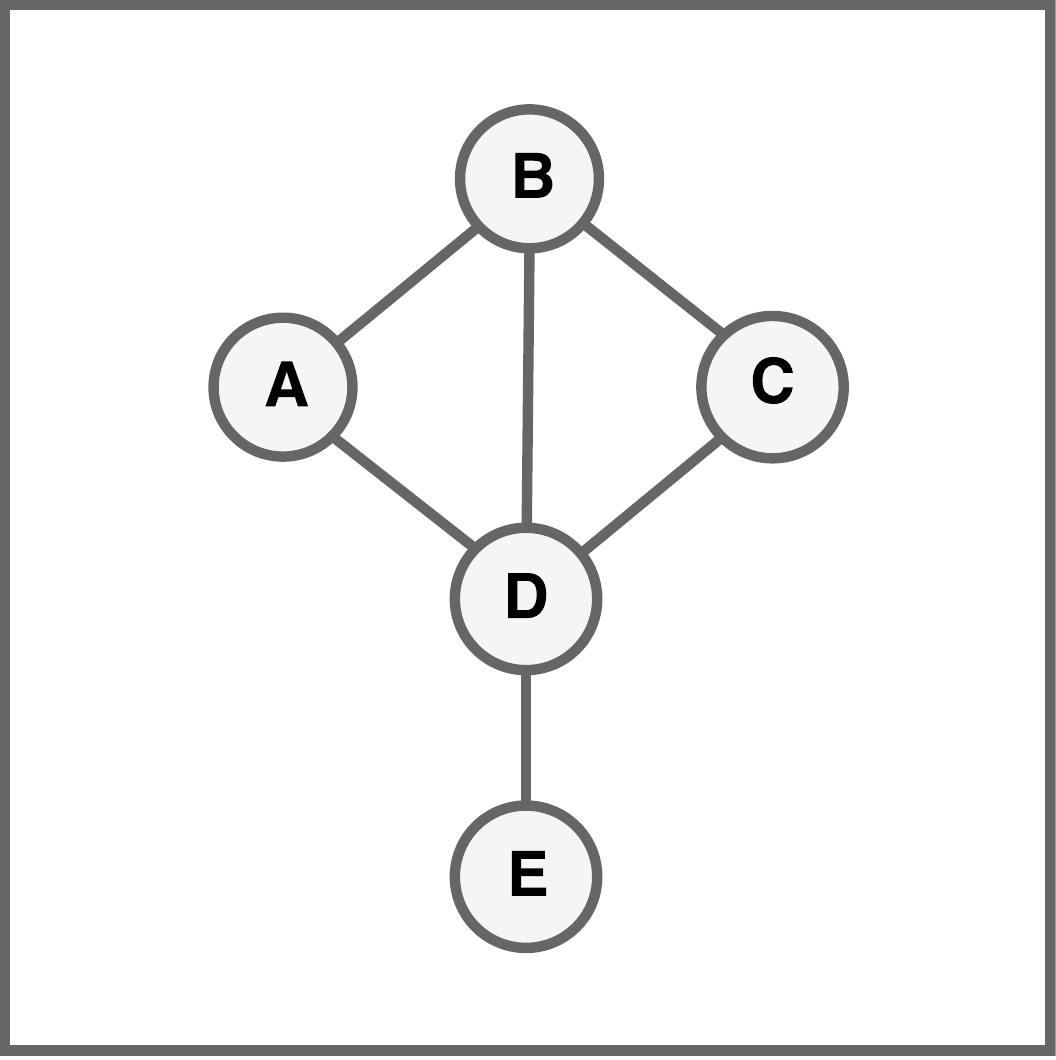}
			\caption{\ac{PO}-graph.}
			\label{fig:pipe1}
		\end{subfigure}&
		\begin{subfigure}[c]{0.28\linewidth}
			\includegraphics[width=\textwidth]{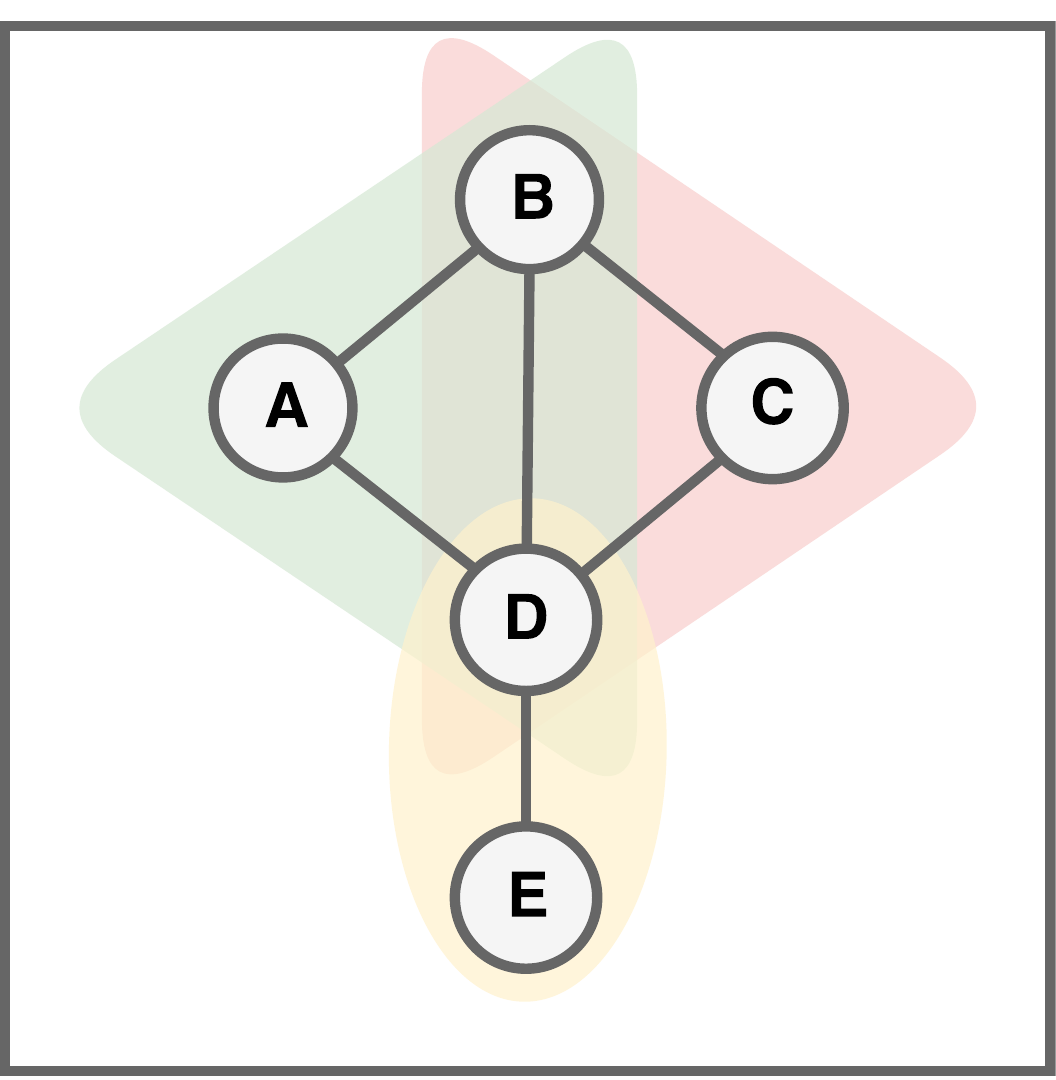}
			\caption{Partitions.}
			\label{fig:pipe2}
		\end{subfigure}\\
		\\
		\multicolumn{2}{l}{
			\begin{subfigure}[l]{0.56\linewidth}
				\includegraphics[width=\textwidth]{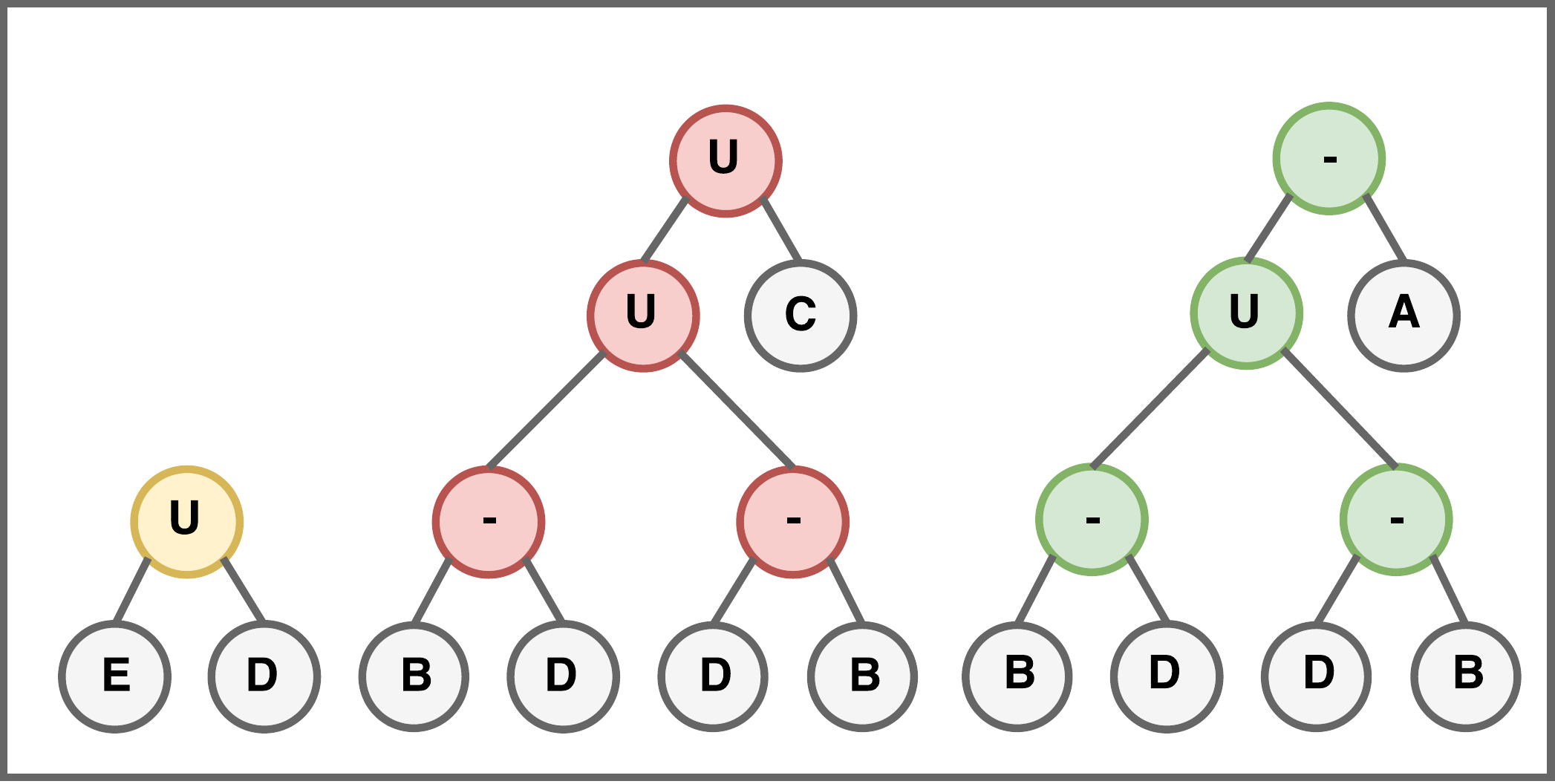}
				\caption{Per-partition trees.}
				\label{fig:pipe3}
			\end{subfigure}
		}&
		\begin{subfigure}[c]{0.28\linewidth}
			\includegraphics[width=\textwidth]{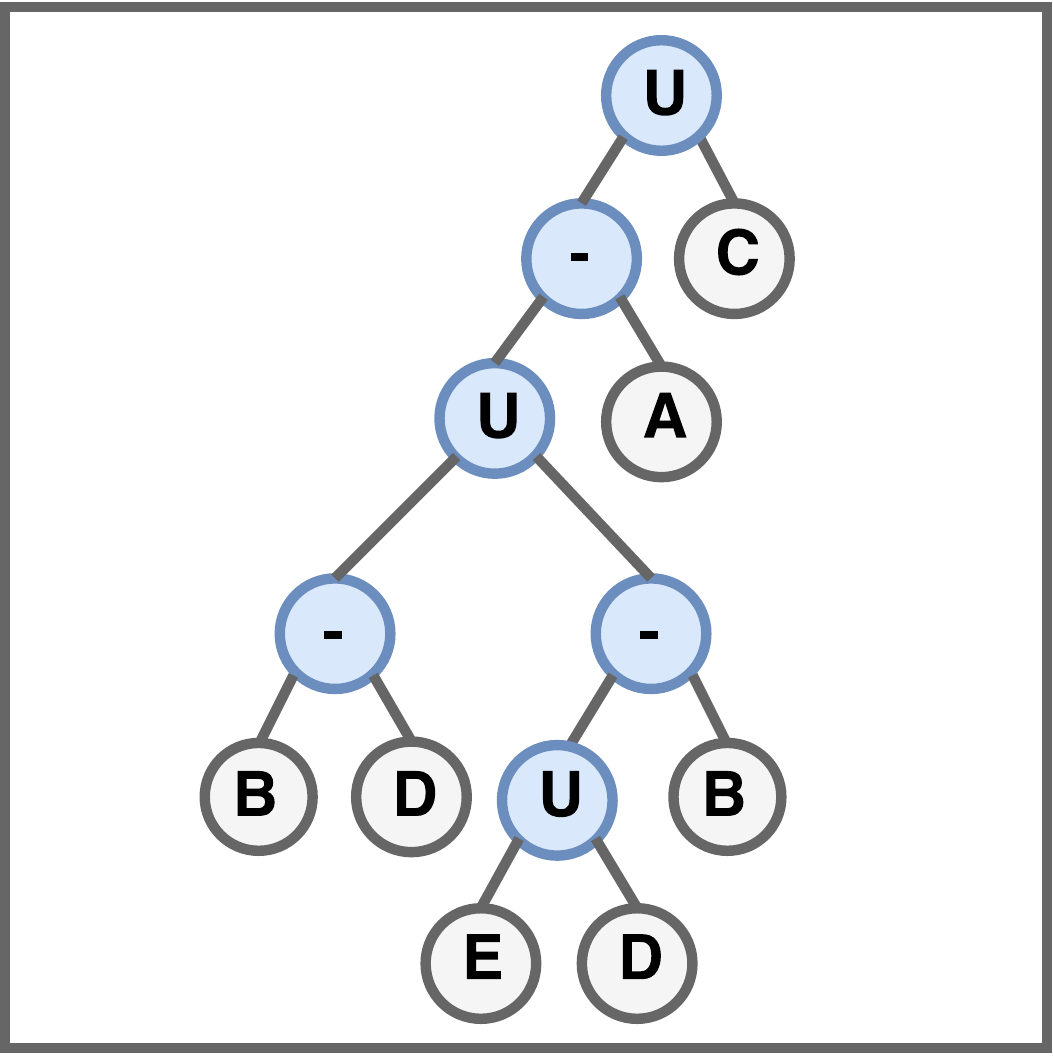}
			\caption{Merged tree.}
			\label{fig:pipe4}
		\end{subfigure}\\	
	\end{tabular}
	\vskip\baselineskip	
	\caption{The search space partitioning pipeline.}\label{fig:pipeline_s}
\end{figure}

The basic idea for accelerating \ac{CSG}-tree extraction is to partition the search space into independent groups of spatially overlapping primitives.
This exploits the fact that primitives that do not overlap are not considered to be operands of a \ac{CSG}-operation.
\ac{CSG}-tree extraction is then conducted on a per-partition level. 
Finally, resulting trees are combined in a subsequent merge step without loss of result quality and correctness. 
\\
An overview of the full \ac{CSG}-tree extraction pipeline is depicted in Fig.~\ref{fig:pipeline_s}. 
Each of the steps is described in detail in the following sub-sections, following the order of execution.
\subsection{Primitive Overlap Graph Generation}
\label{ch:pog}
For expressing spatial relationships between primitives, the \acf{PO}-graph is introduced.
It represents spatial overlap between primitives using an undirected graph $G=(P,O)$, where $P = \{p_1,\dots,p_{n_p}\}$ is the set of $n_p$ primitives as vertices and $O$ is the edge-set that contains $2$-tuples of overlapping primitives $o=(p_i,p_j)$, where $i,j \in \{1,\dots,n_p\}$ with $i \ne j$. 
\\
The \ac{PO}-graph is generated based on the location, orientation and geometric shape of the primitives, see Fig.~\ref{fig:pipe1}.
Complex shapes can be approximated with simpler bounding volumes like \acp{OBB} or the convex hull of the corresponding point-set \cite{preparata1977convex}.
\\
For better scalability, the computational complexity can be reduced from $\mathcal{O}({n_p}^2)$ (overlap check between each primitive and each other primitive) to $\mathcal{O}(n_p\log(n_p))$ using hierarchical space partitioning schemes like e.g. Octrees \cite{meagher1982Octree}.
\subsection{Search Space Partitioning}
\begin{figure}[htb]
	\centering
	\begin{subfigure}[b]{0.25\linewidth}
		\includegraphics[width=\textwidth]{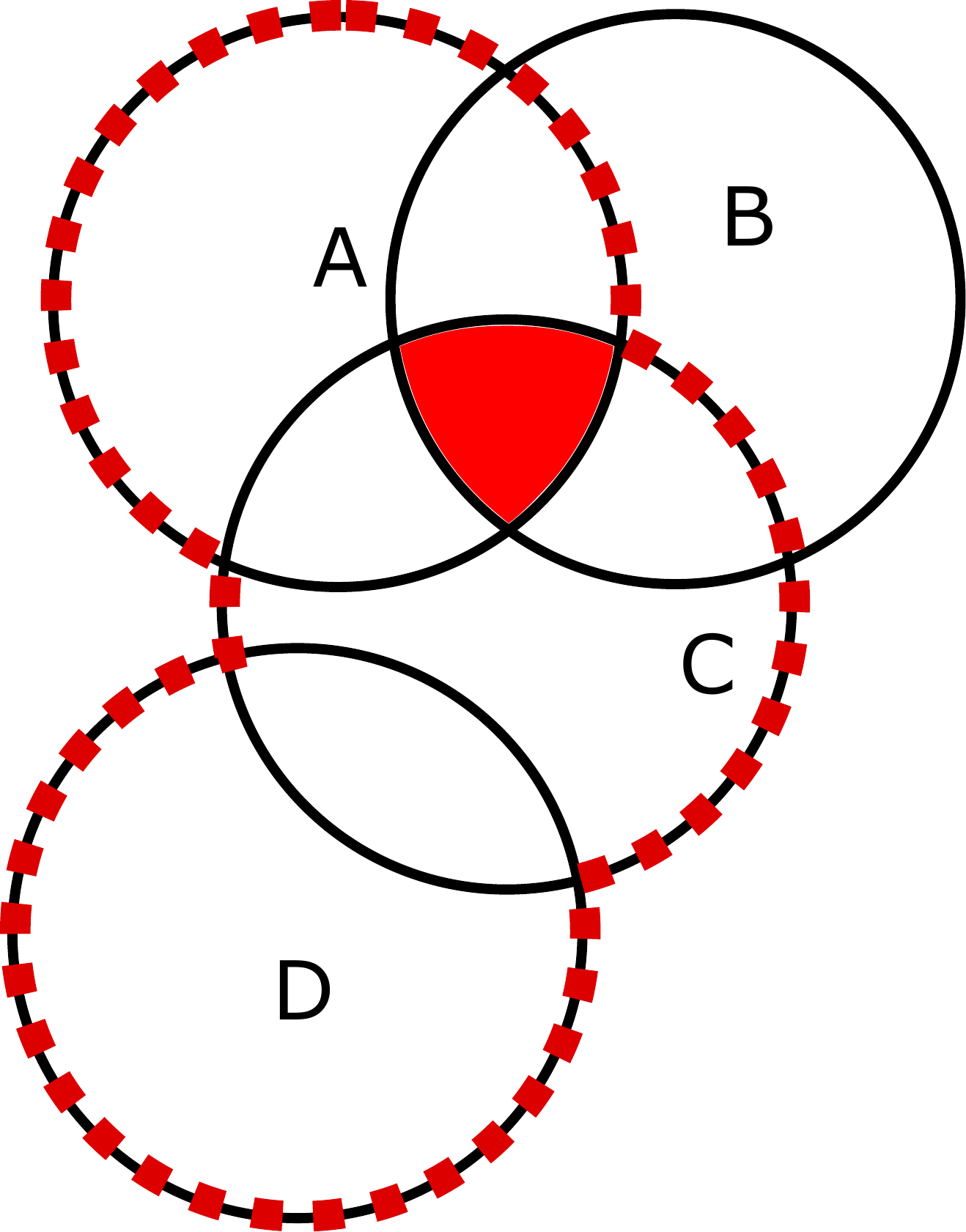}	
		\caption{Incorrect.}
	\end{subfigure}
	~
	\begin{subfigure}[b]{0.25\linewidth}
	\includegraphics[width=\textwidth]{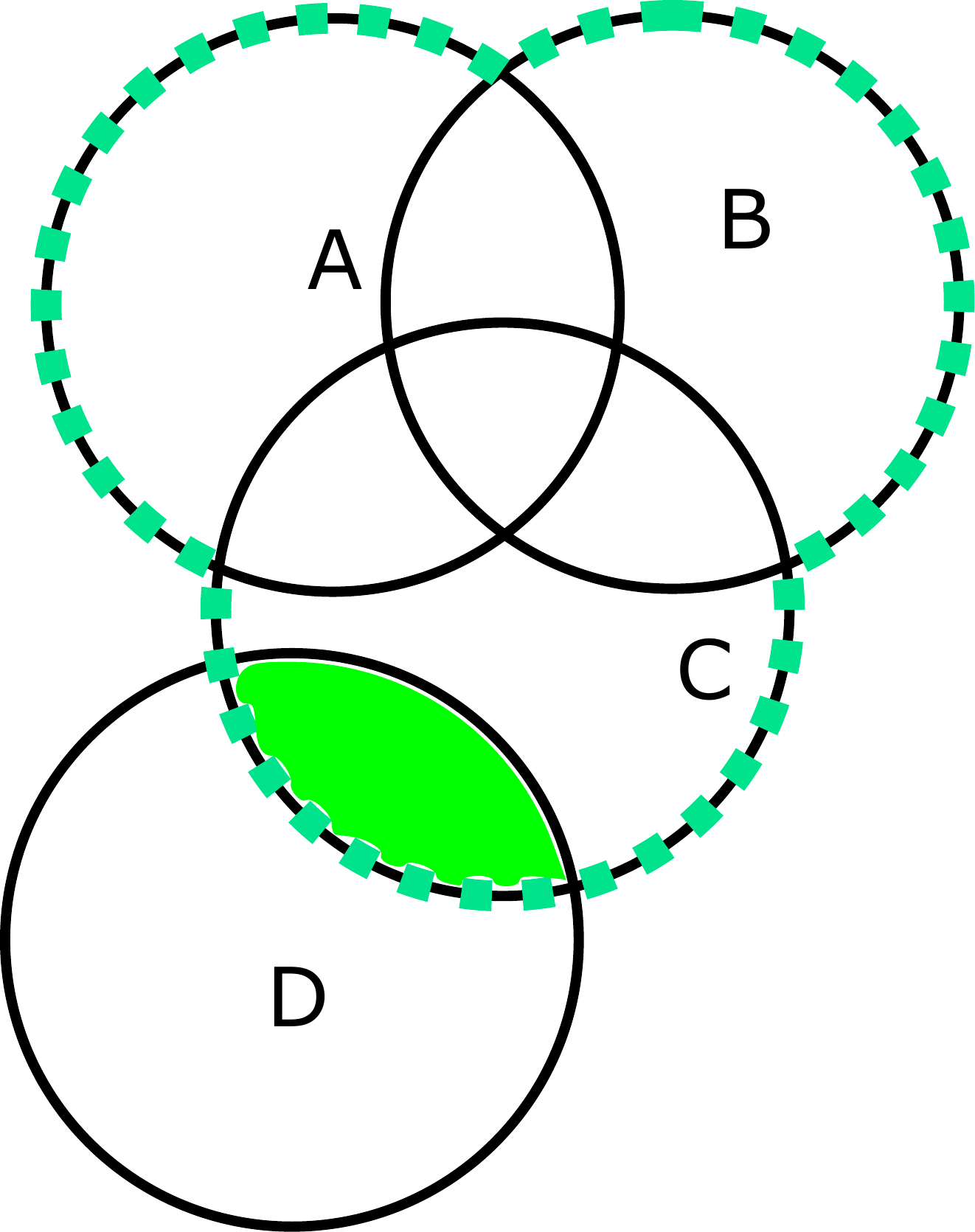}
	\caption{Correct.}
	\end{subfigure}
	\vskip\baselineskip	
	\caption{In (a), per-patition solution parts containing A and C are partially influenced by B (red area) but B is not part of the partition. In (b), D is not part of the partition and influences C only in an area (green) that does not overlap with other partition members.
	Thus, per-partition solutions are not influenced by D. }\label{fig:part}
\end{figure}

With known primitives and their spatial relations given by the \ac{PO}-graph, the goal is now to find independent search space partitions. 
\\
A partition is a set of primitives in which each primitive has an overlap with each other primitive.
In this context, independence means that per-partition solutions are not influenced by the solutions of other partitions.
See Fig.~\ref{fig:part} for explanatory examples. 
\\
The problem of finding all independent search space partitions is equivalent to the problem of finding all maximum complete subgraphs (maximum cliques) in $G$.
For finding the set of maximum cliques in $G$, the \ac{BKA}\cite{bron1973cliques} is employed due to its behavior on random graphs:
It was experimentally shown in \cite{bron1973cliques} that the computational complexity of \ac{BKA} is almost independent of graph size for random graphs.
In a worst case scenario (using Moon-Moser Graphs \cite{moon1965cliques}), computational complexity is proportional to $(3.14)^{\frac{n}{3}}$, where $n$ is the size of the graph.
\\
Note that, if there is only a single partition for a particular \ac{PO}-graph, the search space partitioning method degenerates to standard \ac{GA}-based \ac{CSG}-tree extraction. 
\subsection{Per-Partition \ac{CSG}-Tree Extraction}
\label{ch:ga}
With known partitions, \ac{CSG}-tree extraction is conducted for each partition separately in a divide-and-conquer manner.
A variant of the \ac{GA} described in \cite{fayolle2016evolutionary} is used with the objective function
\begin{equation}
\label{eq:of}
E(t) := \sum_{i=1}^{|S|}\left\{e^{-d_i(t)^2}+e^{-\theta_i(t)^2}\right\}-\alpha \cdot \text{size}(t),
\end{equation}
where $t$ is the tree candidate, $S$ is the point-set corresponding to the partition's primitives and $\text{size}(t)$ is the number of nodes in tree $t$ weighted by a factor $\alpha$.
$d_i(t) = \beta \cdot f_t(s_i)$ is the signed distance between point $s_i$ and the surface defined by tree $t$ weighted by a factor $\beta$.
$\theta_i(t) = \gamma \cdot  \arccos(\nabla \hat{f_t}(s_i) \cdot n_i)$ is the angle between the point normal $n_i$ and the normalized gradient at position $s_i$ weighted by a factor $\gamma$.  
$\alpha, \beta$ and $\gamma$ are user-controlled parameters. 
The first term in Equation \ref{eq:of} (under the sum) estimates how close the surface induced by $t$ matches the point cloud, while the second term penalizes trees with a large number of nodes.
The given objective function has to be maximized for $t$.
\\
Initially, the population $T_0$ is filled with $n_T$ randomly generated trees with a height $\le h_{max}$. 
For the maximum tree height, the approximation  
\begin{equation}
\label{eq:hmax}
h_{max}\approx \sqrt{\pi/2 \cdot n_{pp}\cdot(n_{pp}-1)}
\end{equation}
is used, where $n_{pp}$ is the number of primitives in the partition.
It is based on the average height of binary trees for a given number of internal nodes \cite{flajolet1982TheAH} and achieved good results in all experiments carried out. 
\\
Each \ac{GA} iteration $i$ contains the following steps:
\begin{enumerate}
\item The population of the previous iteration $T_{i-1}$ is ranked according to Equation \ref{eq:of}.
\item The current population is initialized with the $n_b$ best candidates from $T_{i-1}$.
\item As long as $T_i$ has not reached maximum population size $n_T$, two candidates are selected from $T_{i-1}$ via Tournament Selection parameterized with $k_{ts}$ (the size of the set of randomly chosen population members from which the best member is selected) \cite{miller95genetic}.
During crossover, the two selected candidates exchange randomly selected subtrees with a probability of $\gamma_{cr}$.
Then, with a probability of $\gamma_{mu}$, each resulting tree is mutated. 
Either a randomly chosen subtree is replaced with a new randomly generated subtree with a probability of $\mu_{mu}$. Or, with a probability of $1-\mu_{mu}$, the whole tree is replaced with a new randomly generated tree.
\item The termination condition is met if the score of the best \ac{CSG}-tree candidate of an iteration does not improve over $n_{tc}$ iterations.  	 
\end{enumerate}  
The most computationally expensive step in \ac{GA}-based \ac{CSG}-tree recovery is the evaluation of Equation \ref{eq:of} for each element of a candidate-set. 
Since evaluations can be conducted for each candidate independently, parallel processing schemes can be applied efficiently.  
In addition, the solution space partitioning allows for a per-partition parallelization strategy.
Both options were implemented for multi-core processors. Their evaluation is discussed in Section~\ref{sec:eval}.
\subsection{Merge of Per-Partition Trees}
\label{sec:merge}
\begin{figure}[htb]
	\centering
	\begin{tabular}[c]{ccc}
		\multicolumn{2}{c}{
			\begin{subfigure}[c]{0.6\linewidth}
				\includegraphics[width=\linewidth]{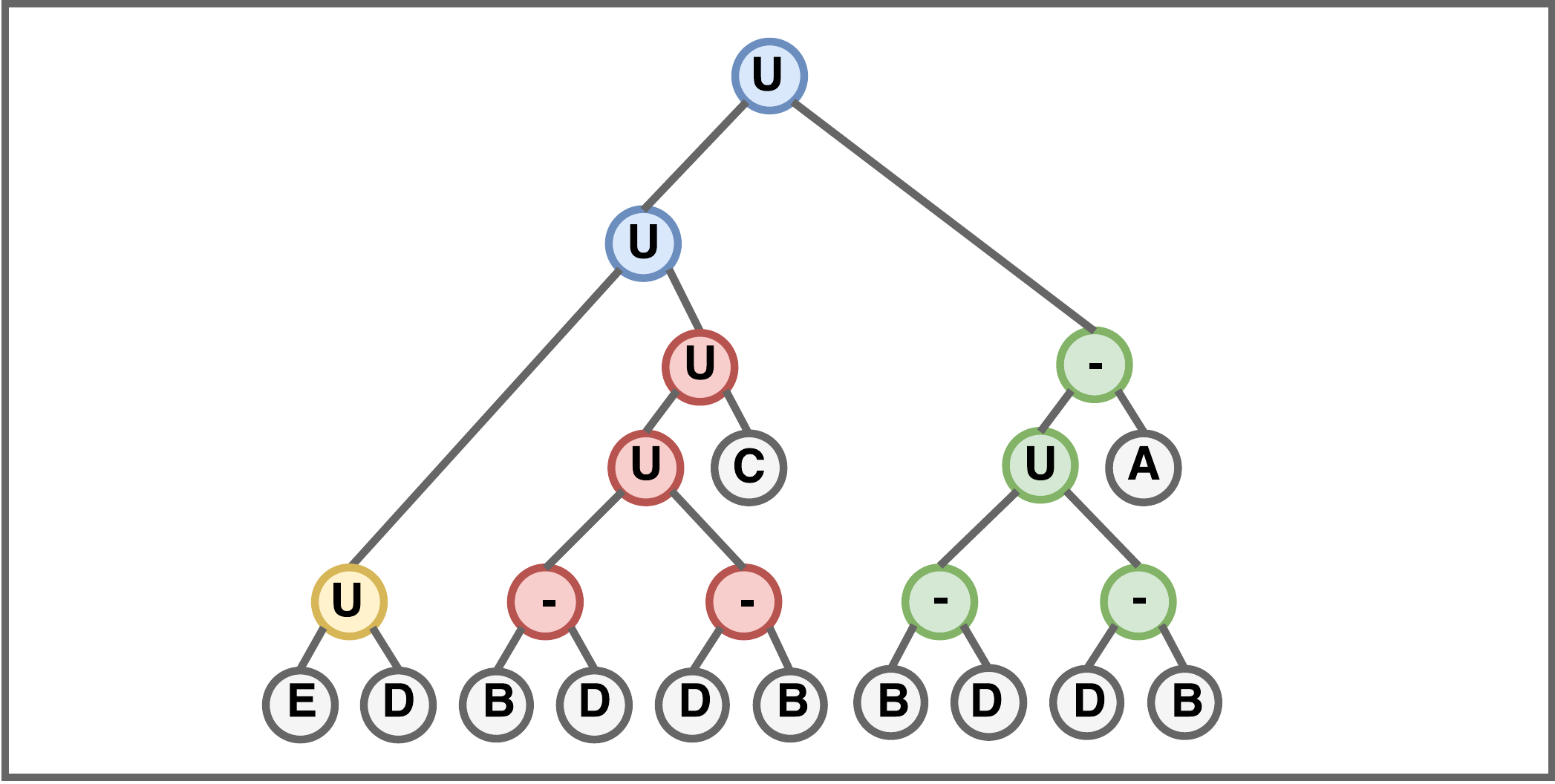}
			\end{subfigure}
		}&
		\begin{subfigure}[c]{0.3\linewidth}
			\includegraphics[width=\linewidth]{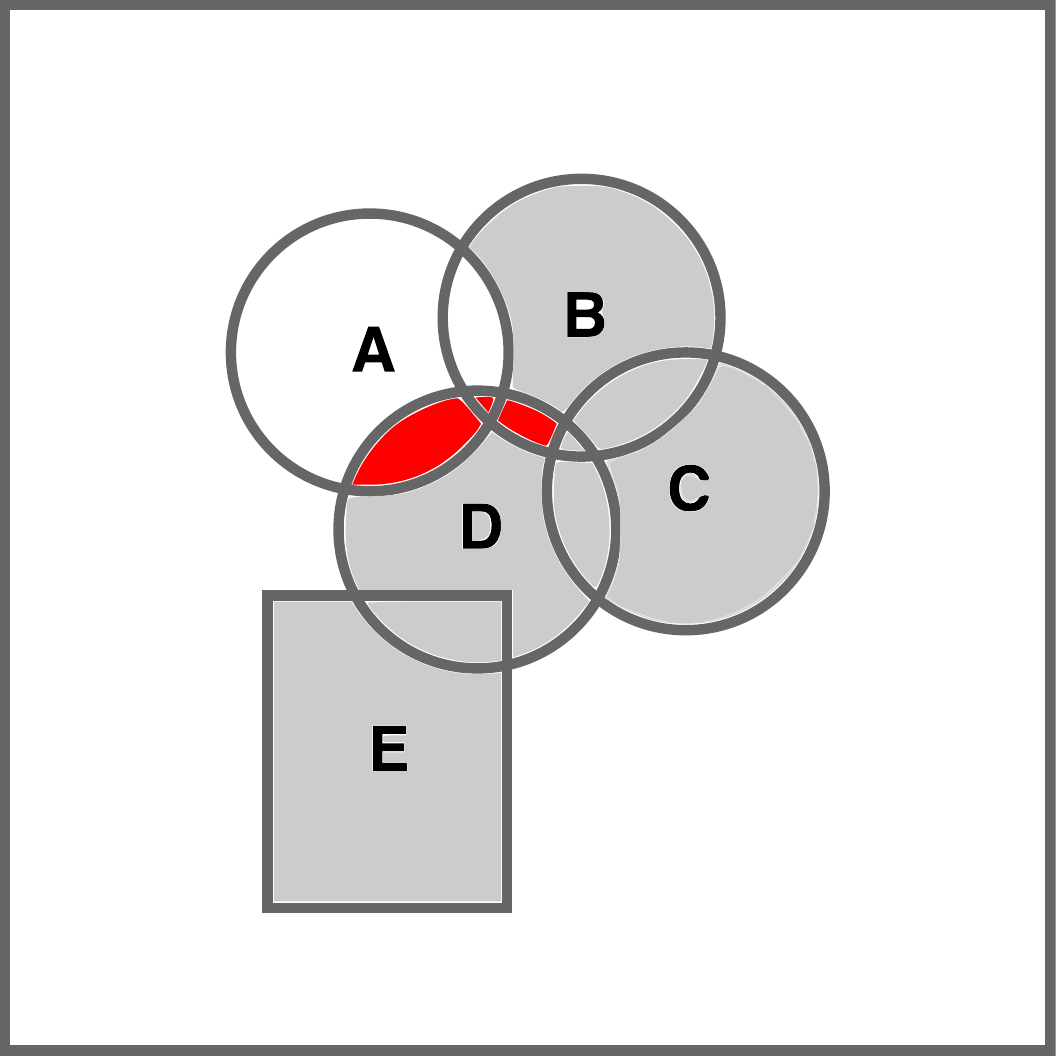}
		\end{subfigure}\\
		\\
		\begin{subfigure}[c]{0.3\linewidth}
			\includegraphics[width=\linewidth]{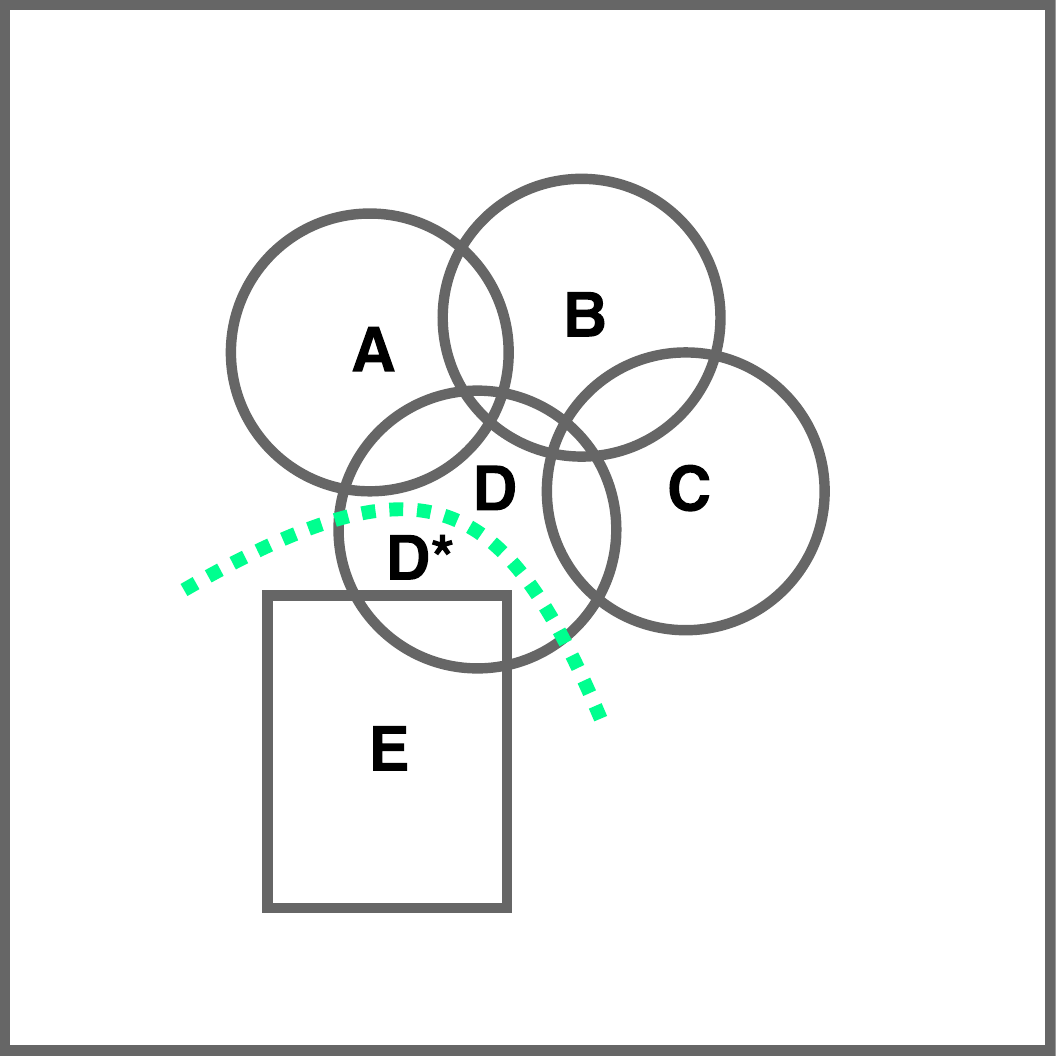}
		\end{subfigure}&
		\multicolumn{2}{c}{
			\begin{subfigure}[c]{0.6\linewidth}
				\includegraphics[width=\linewidth]{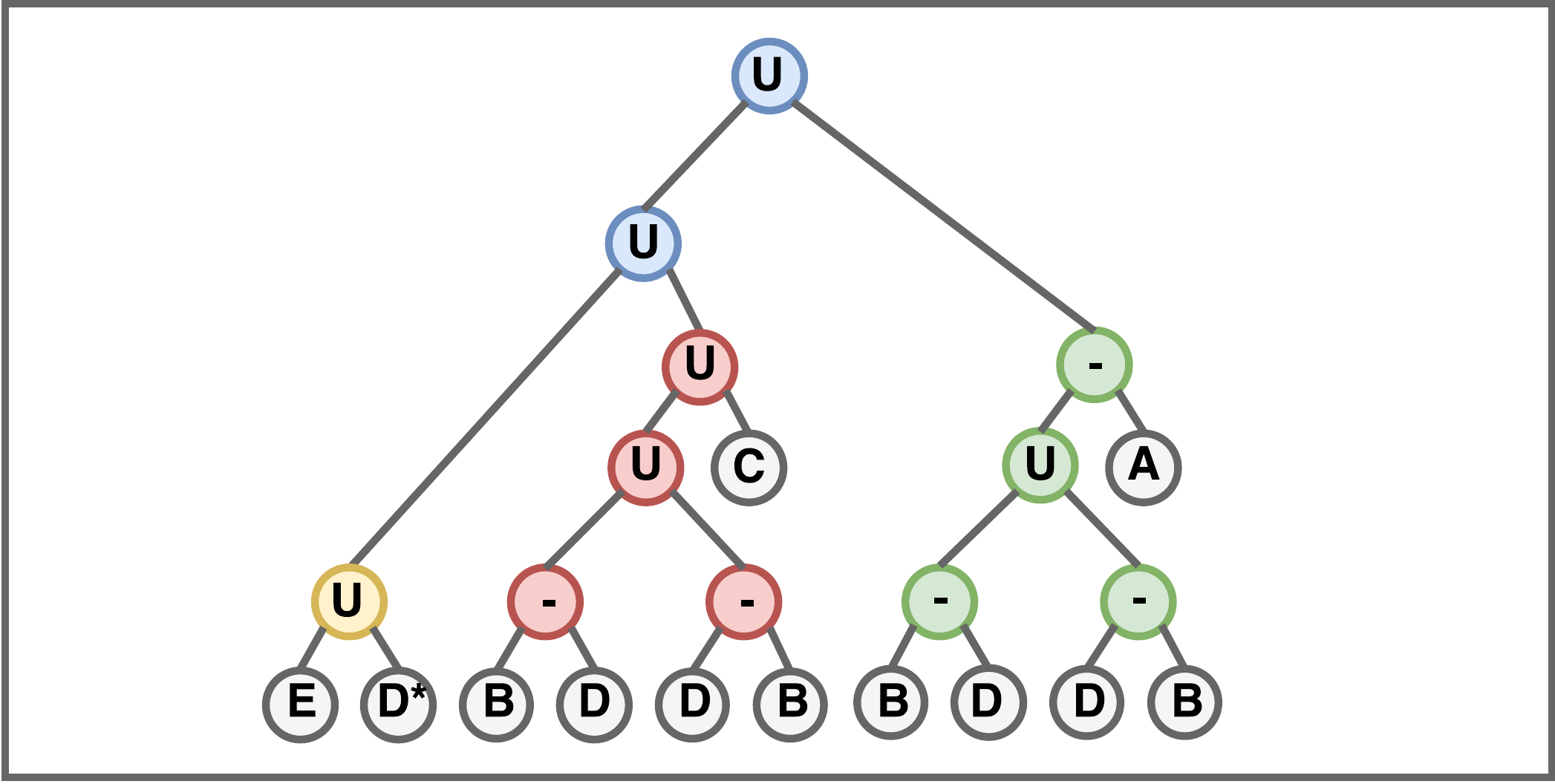}
			\end{subfigure}
		}		
	\end{tabular}
	\vskip\baselineskip
	\caption{Merge strategies. Top: Wrong tree merge using union over all partition trees. 
		Erroneous geometry in red (compare with Fig.~\ref{fig:pipe0}). Bottom: Correct tree merge using union over all partition trees with primitive splitting (green curve).}\label{fig:wmerge}
\end{figure}

\begin{figure}[htb]
	\centering		
	\begin{subfigure}[b]{0.6\linewidth}		
		\includegraphics[width=\textwidth]{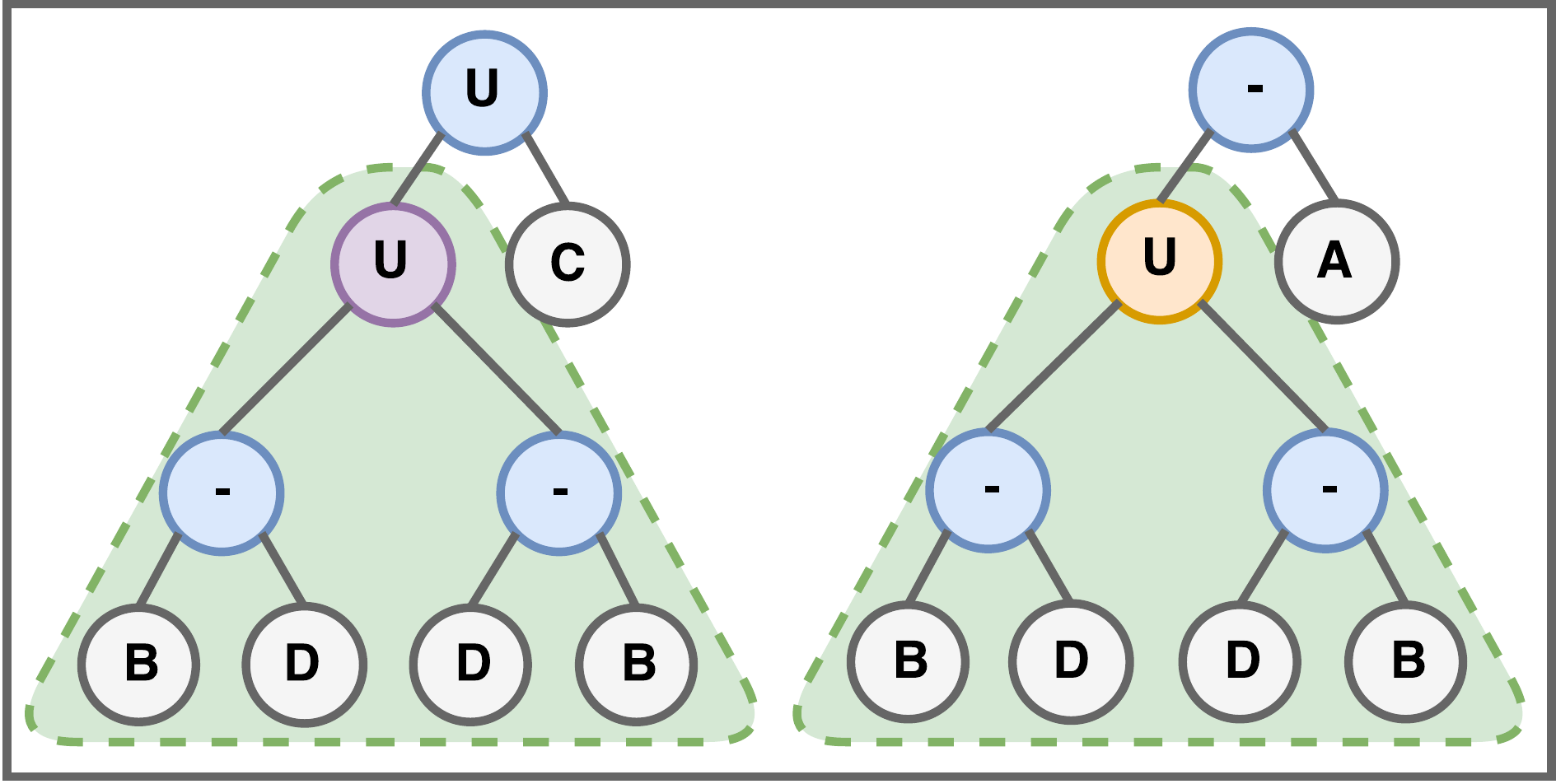}	
		\caption{}			
		\label{fig:subtree_0}	
	\end{subfigure}
	~
	\begin{subfigure}[b]{0.3\linewidth}
		\includegraphics[width=\textwidth]{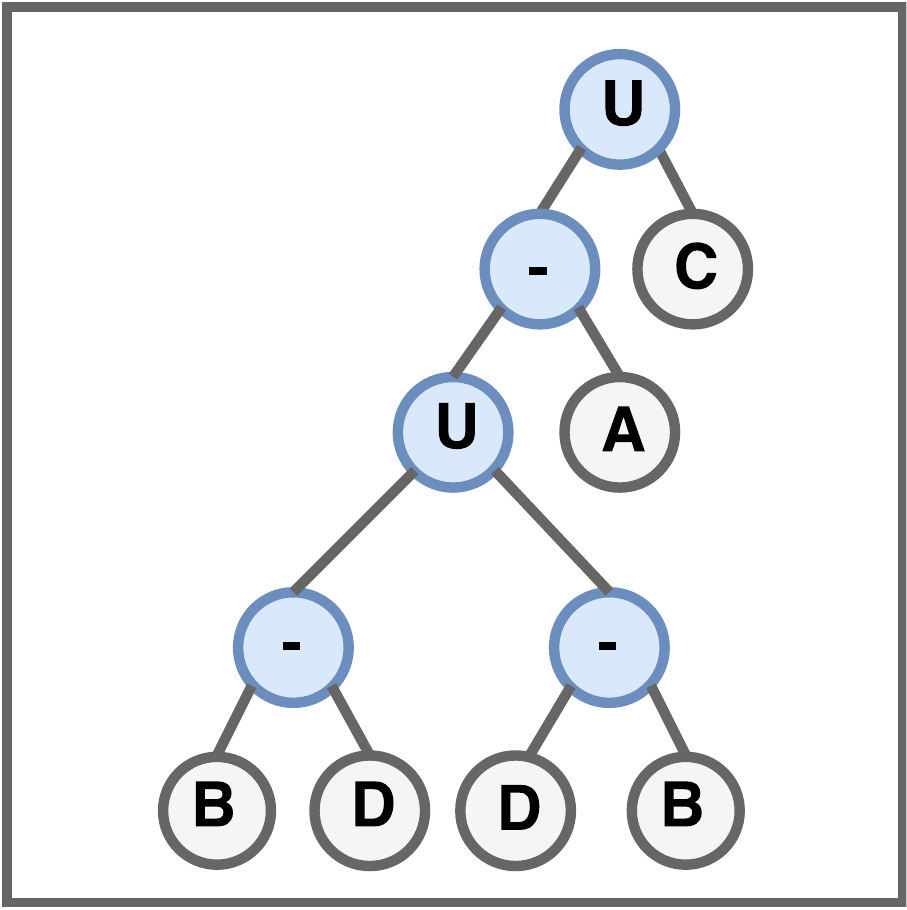}
		\caption{}			
		\label{fig:subtree_1}
	\end{subfigure}	
	\vskip\baselineskip
	\caption{(\protect\subref{fig:subtree_0}) Two merge trees ($t_0$ left, $t_1$ right) with a largest common subtree (green). $N_0$ contains the purple node, $N_1$ the orange node. (\protect\subref{fig:subtree_1}) The merged tree $t_m$.}
\end{figure}

Merging all trees corresponding to partitions into a single tree is not trivial. 
A simple union of all tree root nodes may lead to incorrect results if primitives that are part of multiple cliques are not splitted.
Split operations on arbitrary primitive shapes tend to be complex and should be avoided.
See Fig.~\ref{fig:wmerge} for examples.  
The proposed merge strategy does not need splits but instead tries to merge trees with a subtree in common.
Result correctness is given since no additional operations are introduced and operation order is preserved.
The strategy consists of the following steps:
\begin{enumerate}
	\item All trees are inserted in a list $L$ without any specific order.   
	Extracted trees might contain artefacts affecting their mergeability (e.g., intersections with the same primitive for both operands). 
	For each tree in $L$, artefacts are removed by traversing the tree and replacing found patterns iteratively with their simplifications 
(e.g., replacing $p \cap p$ with $p$).
	The process ends if no more artefacts can be removed.  
	\item Two trees $t_0$ and $t_1$ are removed from the head of $L$, and their largest common subtree $t_{lcs}$ is computed (with a computational complexity of $\mathcal{O}(\max(\text{size}(t_0),\text{size}(t_1)))$).
	The subtree's leaf-set must be a subset of the leaf-sets of both, $t_0$ and $t_1$. 
	The largest common subtree found might exist more than once in both trees.
	Thus, the root nodes of each appearance of the subtree in $t_0$ and $t_1$ are stored in the lists $N_0$ and $N_1$ (see Fig.~\ref{fig:subtree_0}).
	\\	
	If $t_{lcs}$ is empty, $t_1$ is appended to $L$ and a new tree candidate $t_1$ is removed from the head of $L$. 
	In this case, the largest common subtree search is repeated with the new $t_1$.
	\item For each node in $N_0$ and $N_1$, we check if it is a valid merge candidate by traversing the corresponding tree ($t_0$ or $t_1$) from root node to leaves following Algorithm \ref{al:trav}.
	If the node can be reached this way, it is considered a valid merge candidate.
	The node is then replaced by the root of the other tree resulting in a merged tree $t_m$.
	If more than one valid candidate exists, the candidate corresponding to the larger tree is replaced by the root of the smaller tree.
	If both trees are of the same size, the candidate of $t_0$ is chosen (see Fig.~\ref{fig:subtree_1}).
	\\
	If there is no valid merge candidate, the procedure is repeated with the next smaller common subtree in $t_0$ and $t_1$.
	If no other common subtree exists, $t_1$ is replaced by a new tree candidate from the head of $L$.
	Then, the largest common subtree search and its subsequent steps are repeated with the new $t_1$.
	\item The merged tree $t_m$ is prepended to $L$.
	\item The merge process continues until there is only a single node left in $L$.
	Since the model to reconstruct is connected, a pair of mergeable trees exists in each iteration. 
	Thus, the merge process always terminates.
\end{enumerate}
\begin{Ualgorithm}[htb]
	\SetKwProg{myproc}{def}{\string:}{}
        \SetKwIF{If}{ElseIf}{Else}{if}{:}{elif}{else:}{}
        \SetKw{KwTo}{in}\SetKwFor{For}{for}{\string:}{}%
	\SetKwFunction{proc}{isValid}
	\myproc{\proc{curNode, node}}{
		\If{curNode = node}{
			\KwRet {true}
		}
		\If{curNode.nodeType = Operation}{
			\If{curNode.operationType = Difference}{
				\KwRet {\proc{curNode.children[0], node}}
			}
			\ElseIf{curNode.operationType = Union}{
				\For{child $\in$ curNode.children}{
					\If{\proc{child, node}}{
						\KwRet {true}
					}
				}
			}	 	
		}	
		\KwRet {false}
	}
	\nl	\proc{$t$.root, node}
	\caption{Checks if node \textit{node} is a valid merge candidate in tree $t$.}\label{al:trav}
\end{Ualgorithm}
The merge process has an asymptotic computational complexity of $\mathcal{O}(\vert L \vert^2)$ since in the worst case $L$ has to be traversed for each merge.

\section{Evaluation}
\label{sec:eval}
\begin{figure*}
	\centering
	\begin{subfigure}[b]{0.20\linewidth}
		\centering
		\includegraphics[width=\textwidth]{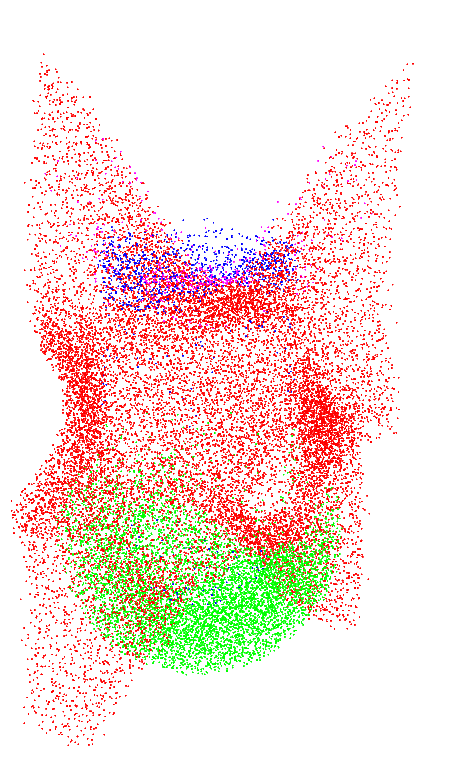}
		\caption{Segmented point-set.}
	\end{subfigure}	 
	\begin{subfigure}[b]{0.25\linewidth}
		\centering
		\includegraphics[width=\textwidth]{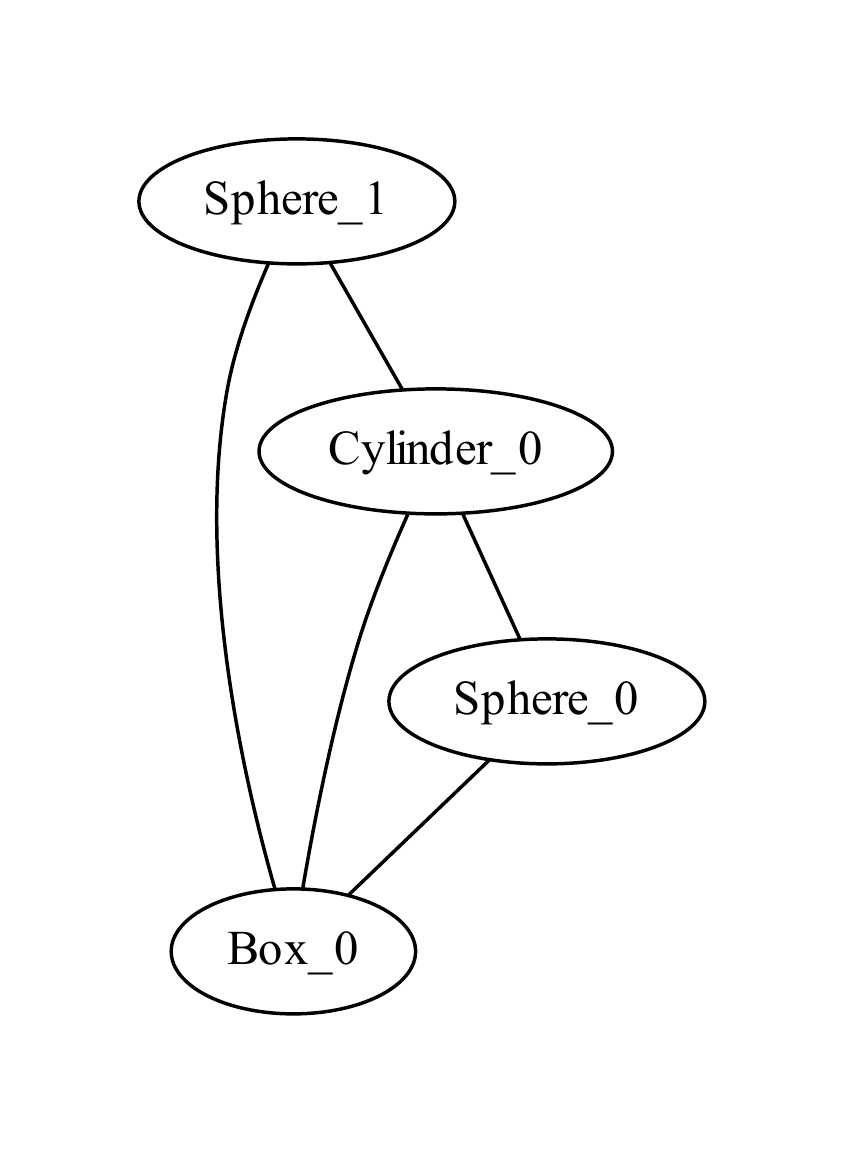}
		\caption{\ac{PO}-graph.}
	\end{subfigure}	
	\begin{subfigure}[b]{0.40\linewidth}
		\centering
		\includegraphics[width=\textwidth]{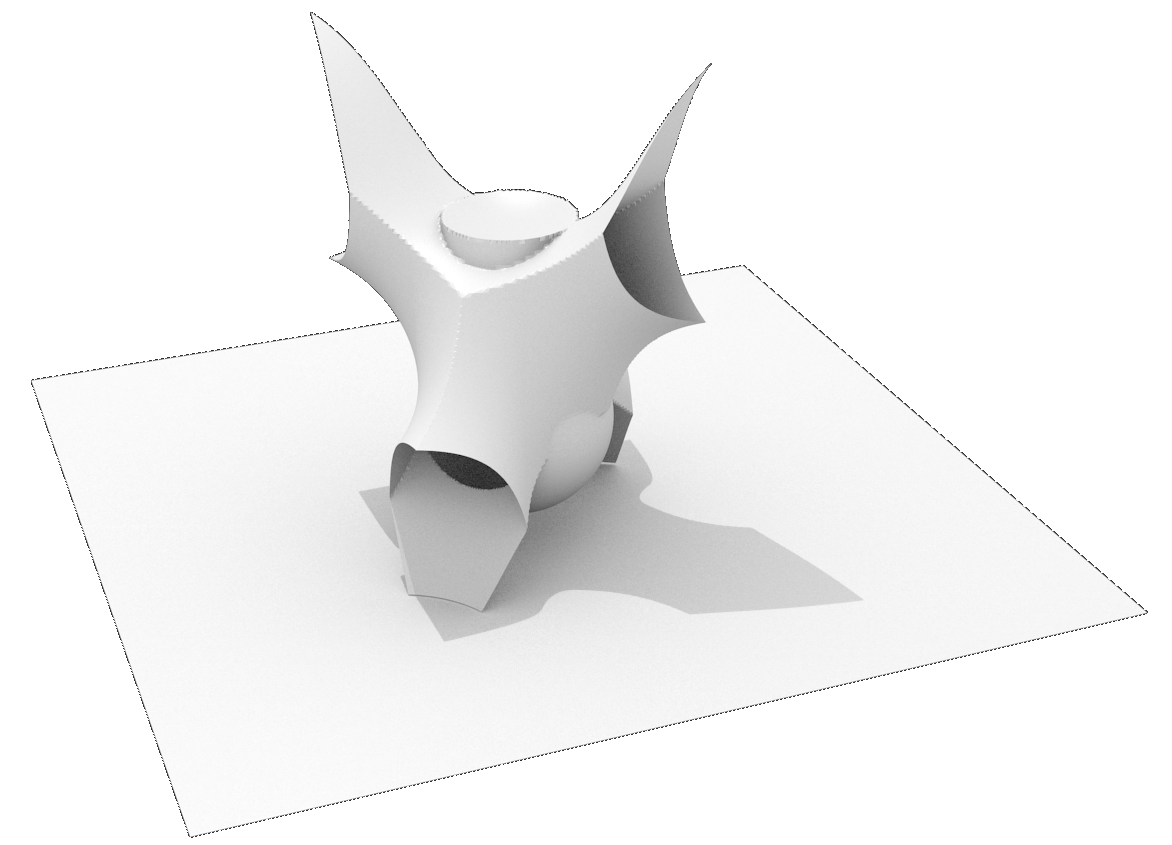}
		\caption{Rendering of resulting model.}
	\end{subfigure}	 
	\\	
	\begin{subfigure}[b]{0.22\linewidth}
		\centering
		\includegraphics[width=\textwidth]{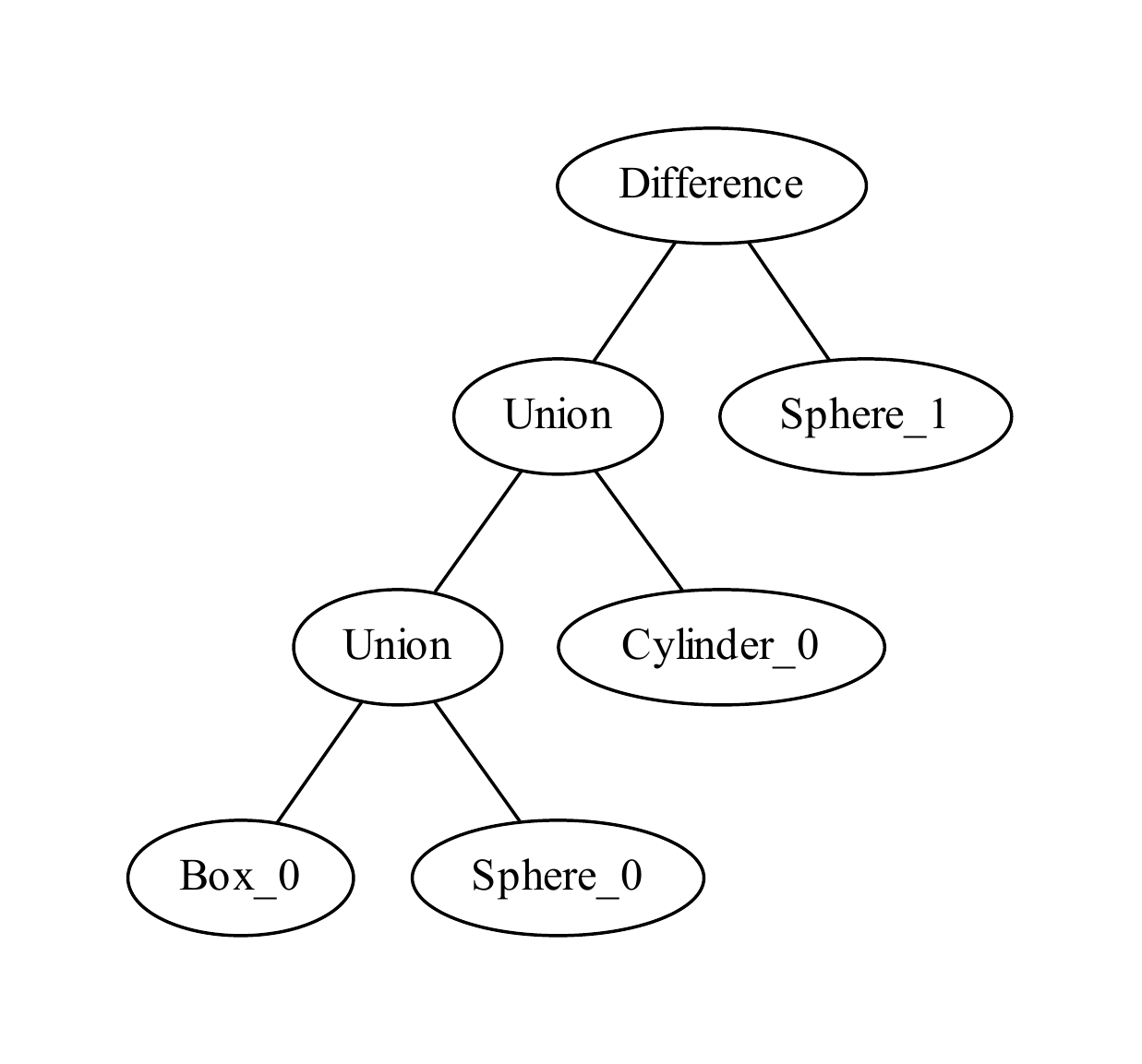}
		\caption{\ac{CSG}-tree ground-truth.}
	\end{subfigure}	 
	\begin{subfigure}[b]{0.38\linewidth}
		\centering
		\includegraphics[width=\textwidth]{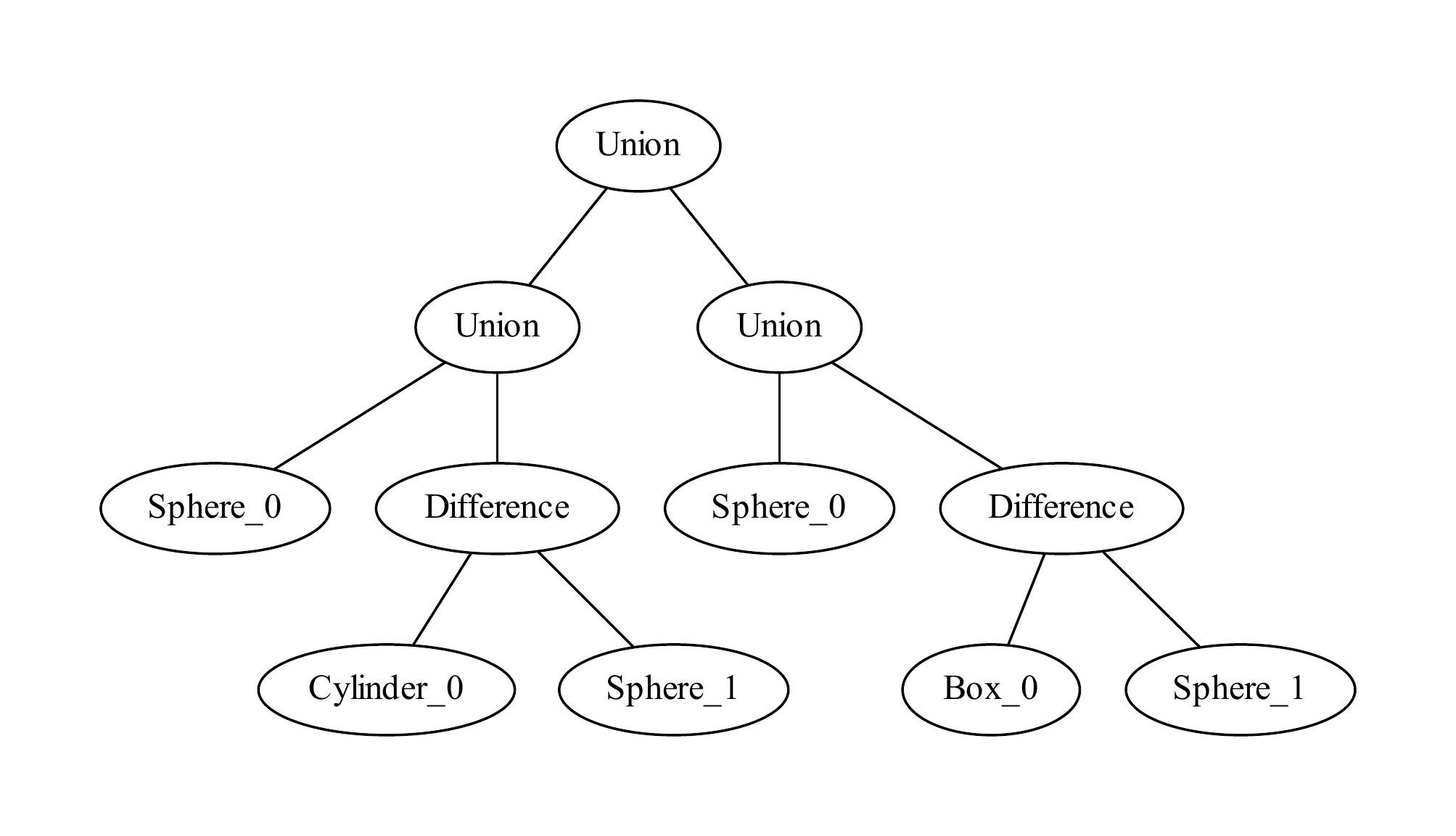}
		\caption{\ac{CSG}-tree from baseline.}
	\end{subfigure}	 
	\begin{subfigure}[b]{0.38\linewidth}
		\centering
		\includegraphics[width=\textwidth]{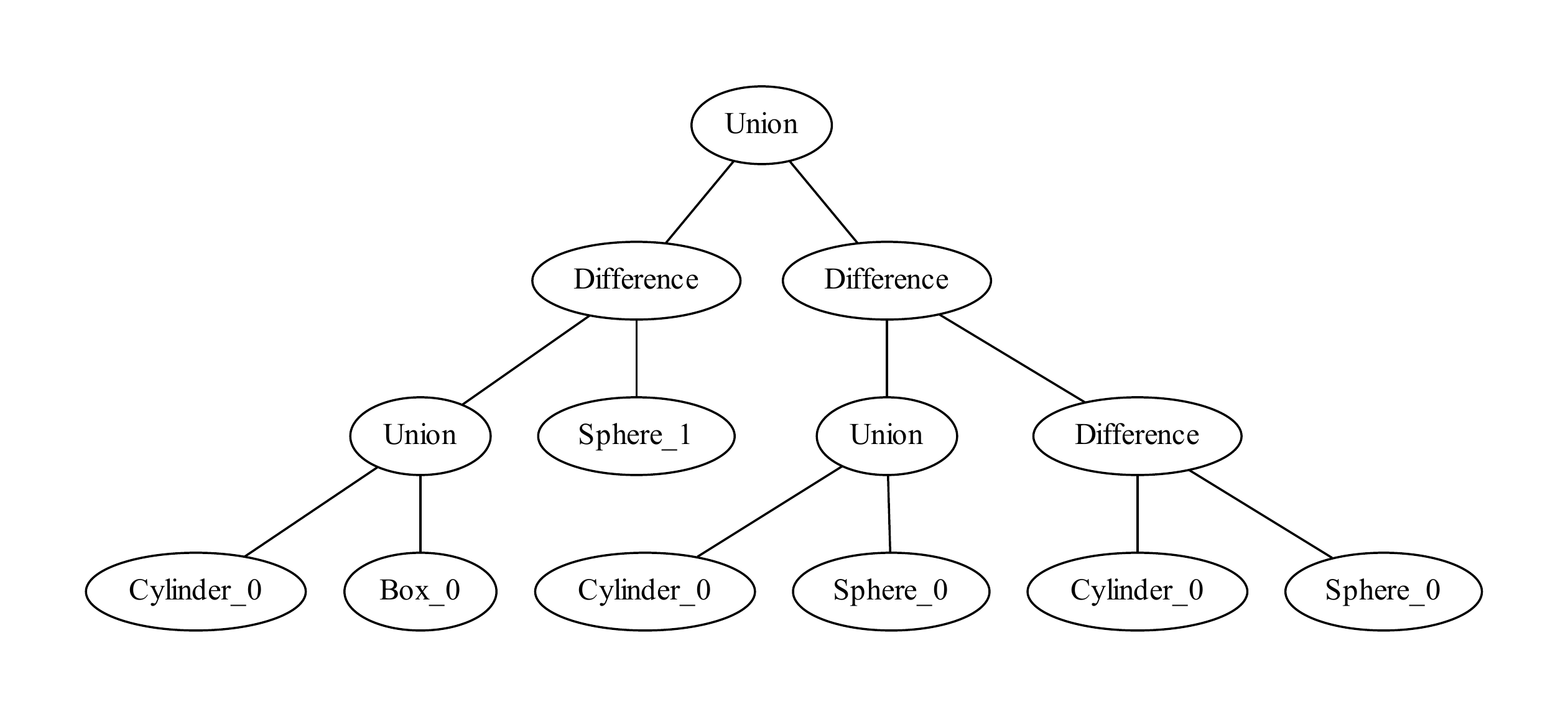}
		\caption{\ac{CSG}-tree from partitioning scheme.}
	\end{subfigure}	
	\vskip\baselineskip	
	\caption{Results of all pipeline steps for model M$1$. The wing-like structure is based on a simple cube whose signed distance function is distorted by a sinusoidal term. This demonstrates the flexibility of the proposed approach in terms of possible model representations.}
	\label{fig:models}
\end{figure*}

\begin{figure*}
	\centering
	\begin{subfigure}[b]{0.30\linewidth}
		\centering
		\includegraphics[width=\textwidth]{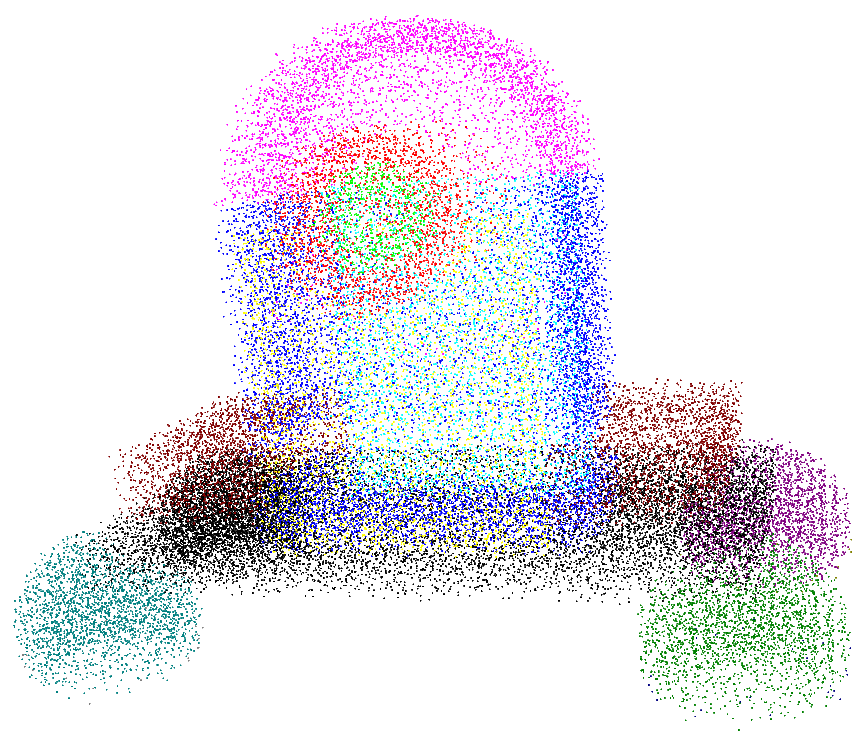}
	\end{subfigure}	 
	\begin{subfigure}[b]{0.3\linewidth}
		\centering
		\includegraphics[width=\textwidth]{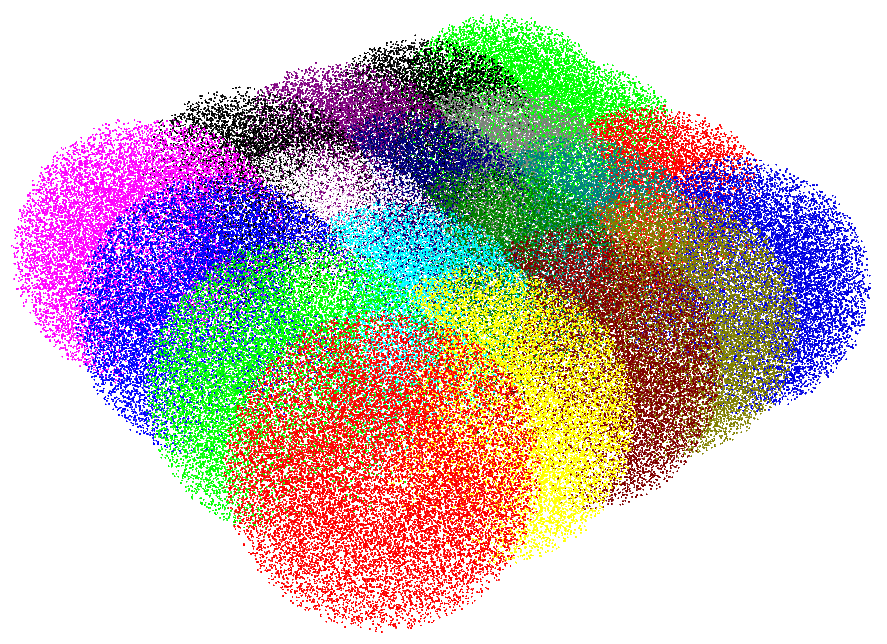}
	\end{subfigure}	
	\begin{subfigure}[b]{0.3\linewidth}
		\centering
		\includegraphics[width=\textwidth]{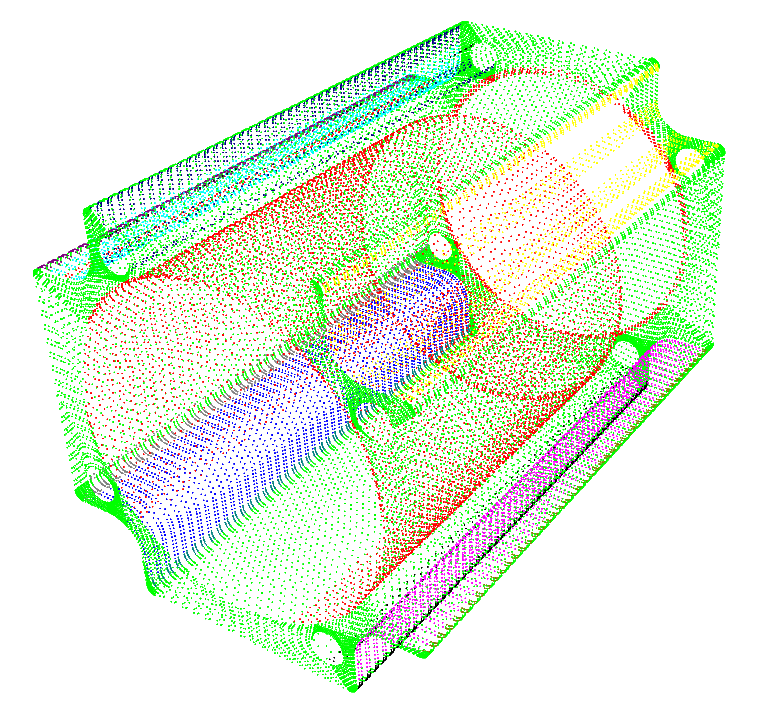}
	\end{subfigure}	 
	\\
	\begin{subfigure}[b]{0.30\linewidth}
		\centering
		\includegraphics[width=\textwidth]{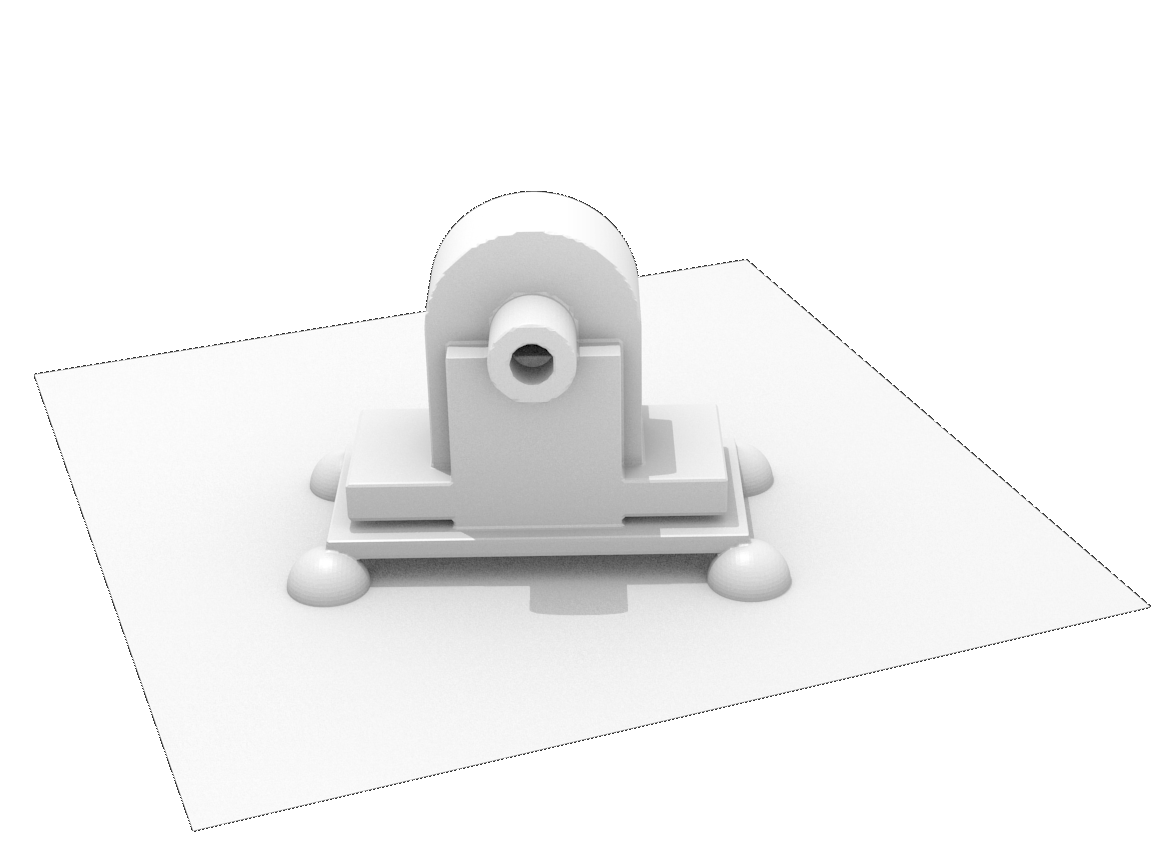}
		\caption{Model M$0$.}
	\end{subfigure}	 
	\begin{subfigure}[b]{0.3\linewidth}
		\centering
		\includegraphics[width=\textwidth]{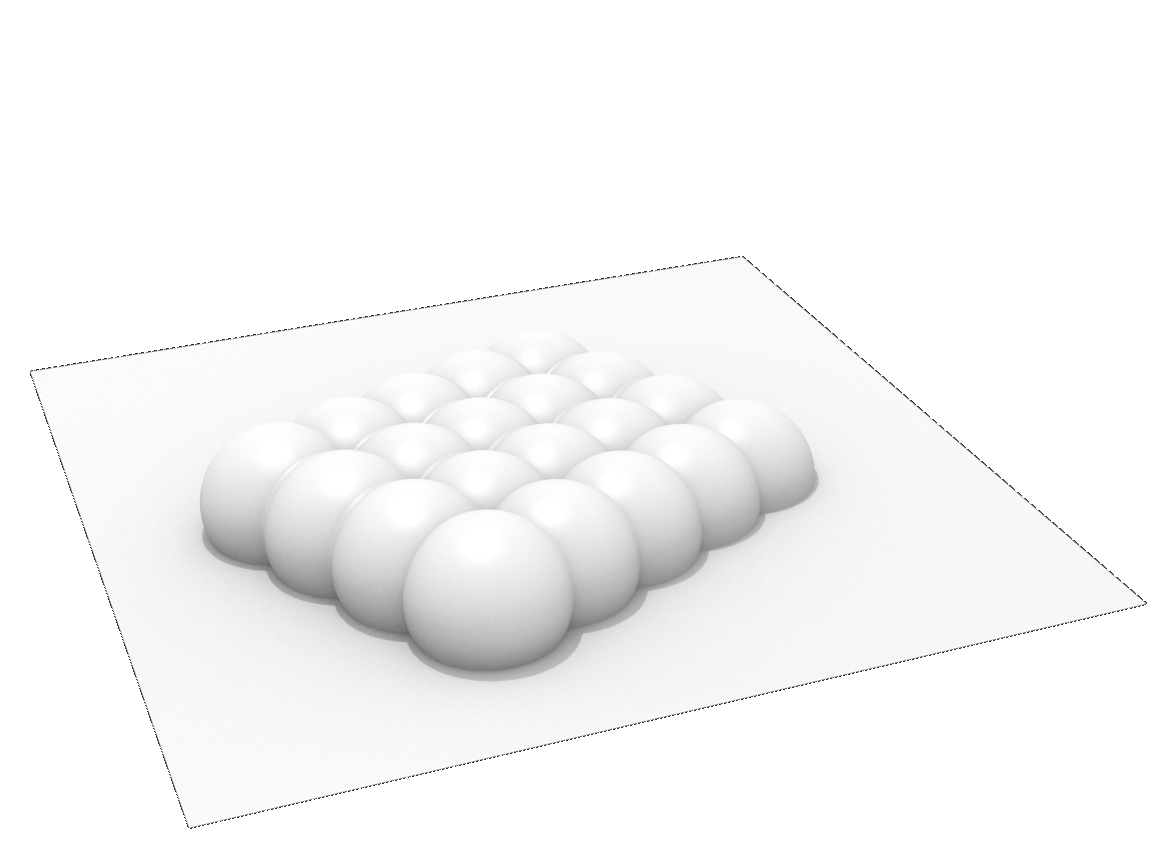}
		\caption{Model M$2$.}
	\end{subfigure}	
	\begin{subfigure}[b]{0.3\linewidth}
		\centering
		\includegraphics[width=\textwidth]{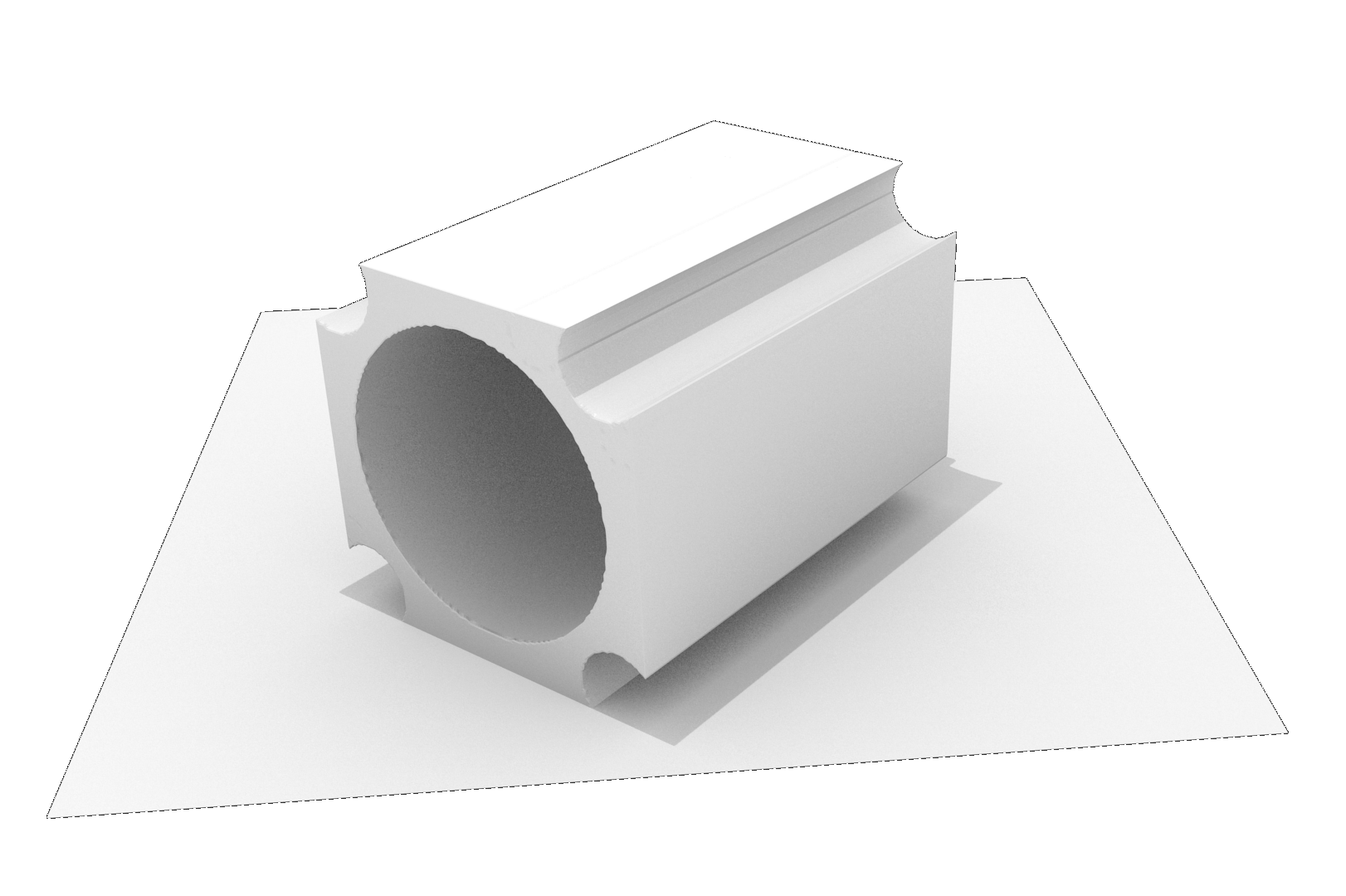}
		\caption{Model M$3$.}
	\end{subfigure}	 
	\vskip\baselineskip	
	\caption{Point clouds and renderings of resulting models M$0$, M$2$ and M$3$.}
	\label{fig:models2}
\end{figure*}

The proposed partitioning scheme has been evaluated on a laptop with quad core CPU and $16$GB of RAM on four different models.
For models M$0$, M$1$ and M$2$, point clouds were generated by sampling
a pre-defined \ac{CSG}-model that served as ground-truth. 
Gaussian noise $(\mu=0.0, \sigma=0.01)$ was added to the points to simulate measurement errors.
Model M$3$ is based on real measurements, and primitive fitting was done with RANSAC \cite{schnabel2007efficient}.
See Fig.~\ref{fig:models} for the intermediate steps results for model M$1$, 
and Fig.~\ref{fig:models2} for point clouds and renderings for models M$0$, M$2$ and M$3$.
Table \ref{tab::models} depicts model details.
\begin{table}[h]
	\centering
	\begin{tabular}{|l|l|l|l|}
	\hline
	 & \textbf{M0} & \textbf{M1} \\
	\hline
	\# Primitives & 17 & 4  \\
	\hline
	\# Points (low) & 11.3k & 9.3k\\
	\hline
	\# Points (high) & 156.4k & 158.4k\\
	\hline
	\# Partitions & (0,8,4,0,1,1) & (0,0,2) \\
	\hline
	& \textbf{M2} & \textbf{M3} \\
	\hline
	\# Primitives & 29 & 18  \\
	\hline
	\# Points (low) & 10.9k & - \\
	\hline
	\# Points (high) & 155.4k & 55.8k \\
	\hline
	\# Partitions & (0,0,0,12) & (0,7,4,1) \\	
	\hline	
	\end{tabular}
	\caption{Details on evaluated models. 'low' and 'high' indicate different sampling rates. Numbers of partitions are depicted per partition size. First position in parantheses indicate number of partitions of size $1$ and so on.}
	\label{tab::models}
\end{table}
\\
The baseline is the \ac{GA} approach proposed in \cite{fayolle2016evolutionary} and described in Section~\ref{ch:ga}. 
The parameter-set used for both, baseline and partitioning scheme, is listed in Table \ref{tab:gaparams}.
The following combinations were evaluated:
\begin{itemize}
	\item \textbf{Baseline:} Single-threaded (BST), multi-threaded \ac{GA} (BMTGA).
	\item \textbf{Search Space Partitioning:} Single-threaded (SST), per-partition multi-threaded (SMTP) multi-threaded \ac{GA} (SMTGA), per-partition and \ac{GA} multi-threaded (SMTPGA) combined.
\end{itemize}
\begin{table}[h]
	\centering
	\begin{tabular}{|l|l|}
		\hline
		\textbf{Parameter Name} & \textbf{Value}  \\
		\hline
		Population size $n_T$ & 150 \\
		\hline
		\# Best parents $n_b$ & 2 \\
		\hline
		Crossover probability $\gamma_{cr}$& 0.3 \\
		\hline
		Mutation probability $\gamma_{mu}$& 0.3 \\
		\hline
		Subtree replacement probability $\mu_{mu}$& 0.5 \\
		\hline
		Tournament selection parameter $k_{ts}$ & 2\\
		\hline
		Tree size weight $\alpha$& $\log(\text{\#pts.})$\\
		\hline
		Distance weight $\beta$& $100.0$ \\
		\hline
		Angle weight $\gamma$& $18.0/\pi$ \\
		\hline 
		\# Iterations w/o quality increase $n_{tc}$ & 10 \\
		\hline 
		Maximum tree height $h_{max}$ & see Eq.~\ref{eq:hmax} \\
		\hline 
	\end{tabular}
	\caption{Parameters for the baseline and search space partitioning approach.}
	\label{tab:gaparams}
\end{table}   
\subsection{Computation Times}  
\begin{figure}[htb]
	\centering
	\includegraphics[width=0.45\textwidth]{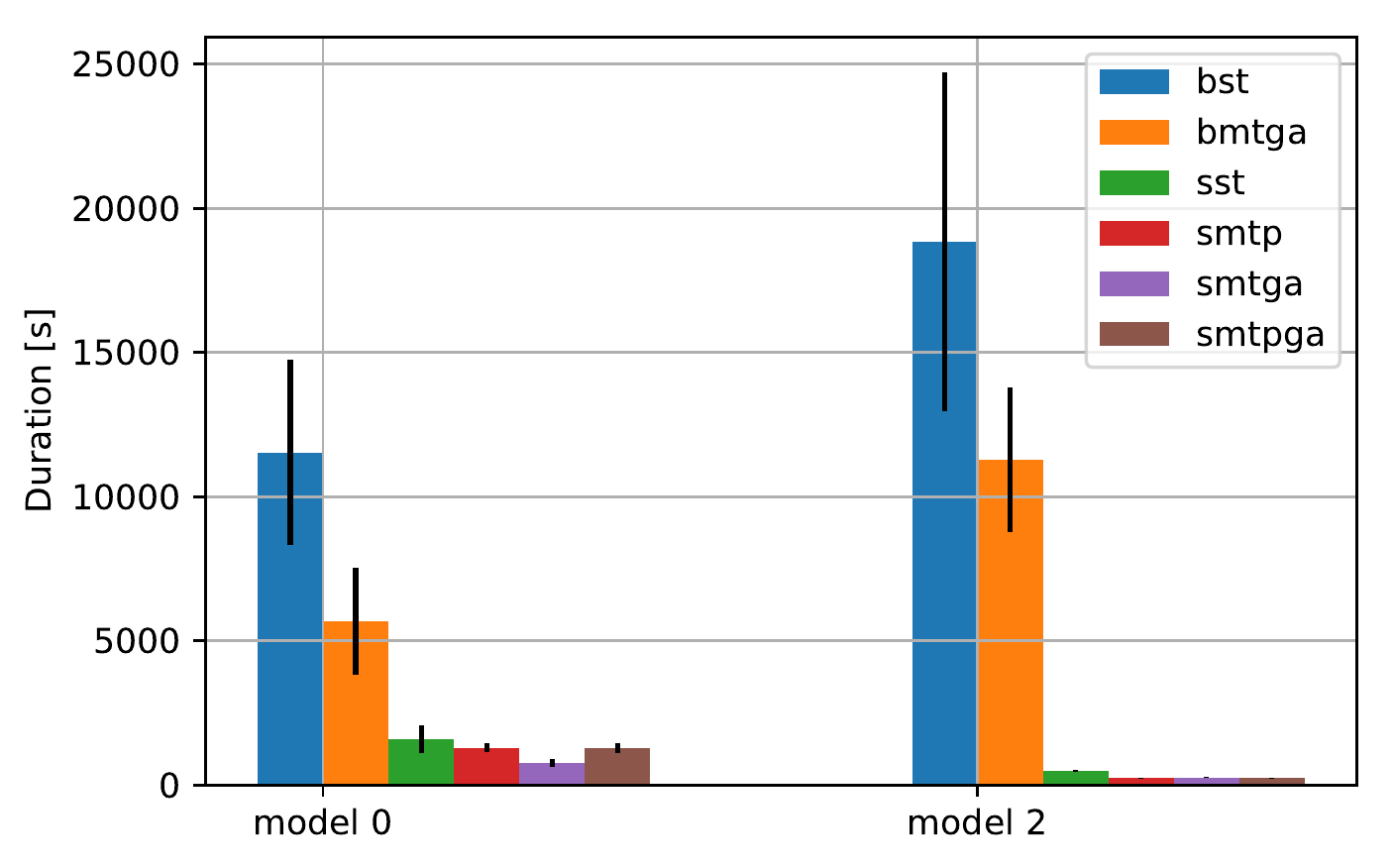}
	\caption{Timings for all approach combinations and models M$0$ and M$2$ with high-detail sampling (black lines: standard deviations).}
	\label{fig:graph1}
\end{figure}

\begin{figure}[htb]
	\centering
	\includegraphics[width=0.45\textwidth]{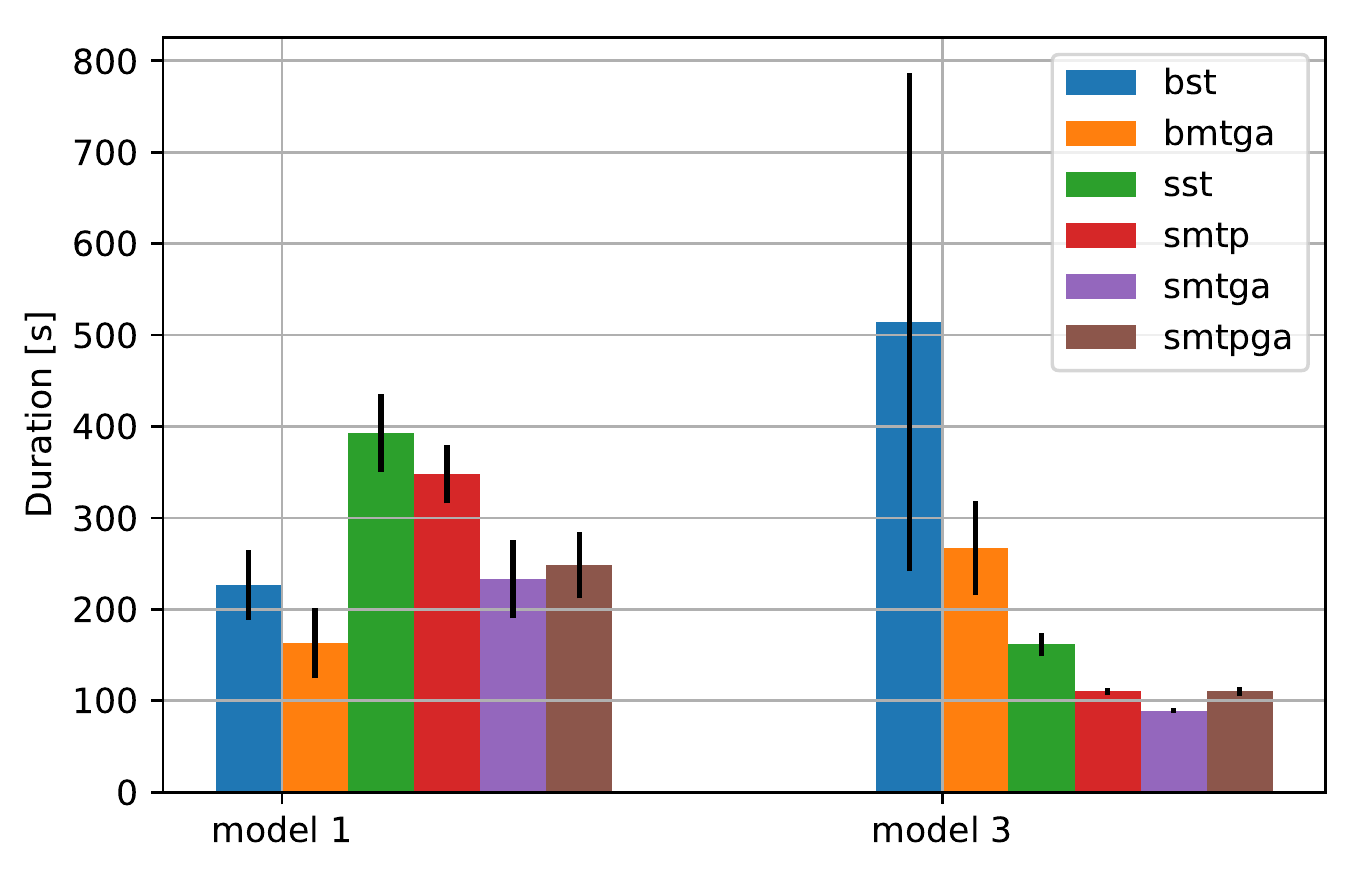}
	\caption{Timings for all approach combinations and models M$1$ and M$3$ with high-detail sampling (black lines: standard deviations).}
	\label{fig:graph4}
\end{figure}

Timings for baseline and search space partitioning variants were measured for all models with high- and low-detail sampling (except for model M$3$ for which only a single point cloud exists).
Measurements vary significantly for the same benchmark setting due to the stochastic behavior of \ac{GA}-based methods. 
In order to deal with this variance, each experiment was repeated $5$ times.
\\
In the following, timing results for all methods in combination with high-detail sampling are discussed.
See Fig.~\ref{fig:graph1} and \ref{fig:graph4} for an overview of the results. 
For model M$0$, SMTGA is the fastest method. 
It outperforms the baseline by a factor of $15.3$ (single-threaded, BST) and $7.5$ (multi-threaded, BMTGA) on average.
For model M$1$, search space partitioning performs worse than baseline: 
The fastest baseline method (BMTGA) is on average $1.4$ times faster than the best-performing search space partitioning variant (SMTGA).
This can be explained by the relatively small number of primitives ($4$) and partitions ($2$) in model M$1$, which reduces the need for partitioning.
For model M$2$, single-threaded partitioning is $38.3$ times faster than single-threaded baseline and multi-threaded partitioning variants are between $43.4$ and $46.6$ times faster than multi-threaded baseline.  
The considerable difference is due to the relatively high number of partitions ($12$) and their equally distributed size (all contain $4$ primitives).
For model M$3$, SMTGA is again the fastest method. 
Compared to multi-threaded baseline it is $3.0$ times faster on average.
\\
Search space partitioning with \ac{GA} parallelization (SMTGA) is in general faster than their per-partition counterparts (SMTP, SMTPGA) for all models.
This is due to the granularity and regularity of the parallelization: 
For SMTGA, the task of ranking a population can be splitted into $n_T$ parts, with each part having similar execution times.
For per-partition variants, granularity is determined by the (potentially lower) number of partitions, and per-partition execution times may vary significantly depending on partition sizes. 
\\
Results for per-partition variants do not show timings for the different pipeline steps since in all experiments, per-partition \ac{CSG}-tree extraction is by far the most dominant factor.
Timings for \ac{PO}-graph generation, search space partitioning and tree merge make less than $1\permil$ of the total runtime.
\subsection{Tree Sizes and Depths}  
\begin{figure}[htb]
	\centering
	\includegraphics[width=0.45\textwidth]{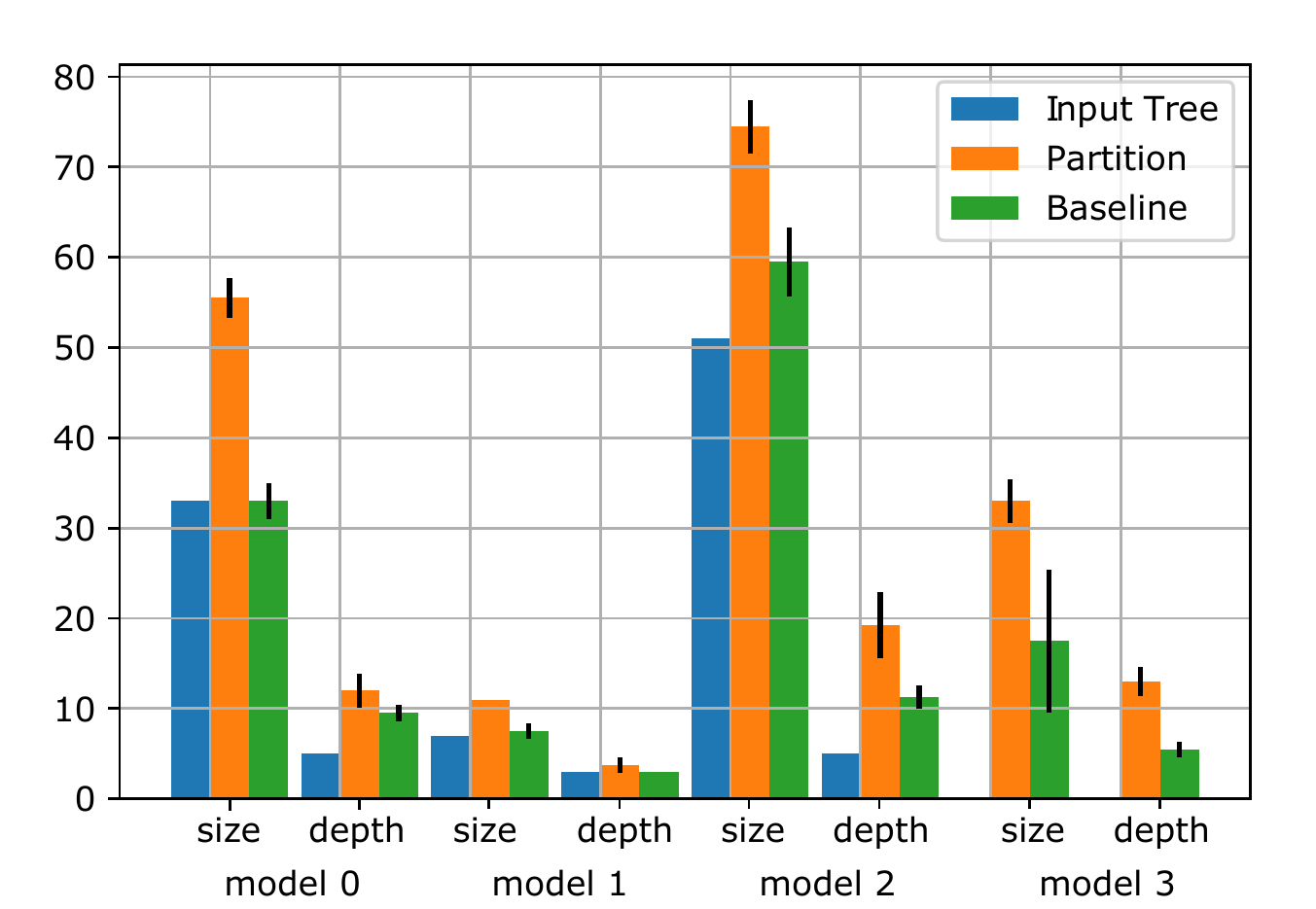}
	\caption{Average tree size and depth for baseline and partitioning methods (black lines: standard deviations).}
	\label{fig:graph2}
\end{figure}

Fig.~\ref{fig:graph2} contains average depths and sizes of resulting trees for baseline and partitioning variants.
For the latter, tree depths have increased by $25.0\%$ (model M$1$) to $285.0\%$ (model M$2$) compared to the input tree, while for baseline approaches, an increase of $0.0\%$ (model M$1$) to $125.0\%$ (model M$2$) is visible.
Tree sizes show similar behavior:
Partitioning variants produce $46.1\%$ (model M$2$) to $68.2\%$ (model M$0$) larger trees, while baseline approaches increase tree size by only $0.0\%$ (model M$0$) to $16.7\%$ (model M$2$).
Comparing tree sizes between partitioning and baseline approaches directly reveals that the former results in $25.2\%$ (model M$2$) to $88.6\%$ (model M$3$) larger trees.
\\
This adverse behavior shown by partitioning variants is due to the final merge step:
In each iteration, the two trees that are close to each other in the tree list and have a common subtree of at least size $1$ are merged instead of the two trees with the largest common subtree of all tree pairs in the merge list. 
Since the focus of this work is on performance, this is acceptable.
In addition, the tree optimization strategy described in Section \ref{sec:merge} (step $1$) was also applied to baseline results for better comparability, which has positive impact on resulting tree depths and sizes.
\subsection{Scalability with Respect to Point Cloud Size}  
\begin{figure}[htb]
	\centering
	\includegraphics[width=0.47\textwidth]{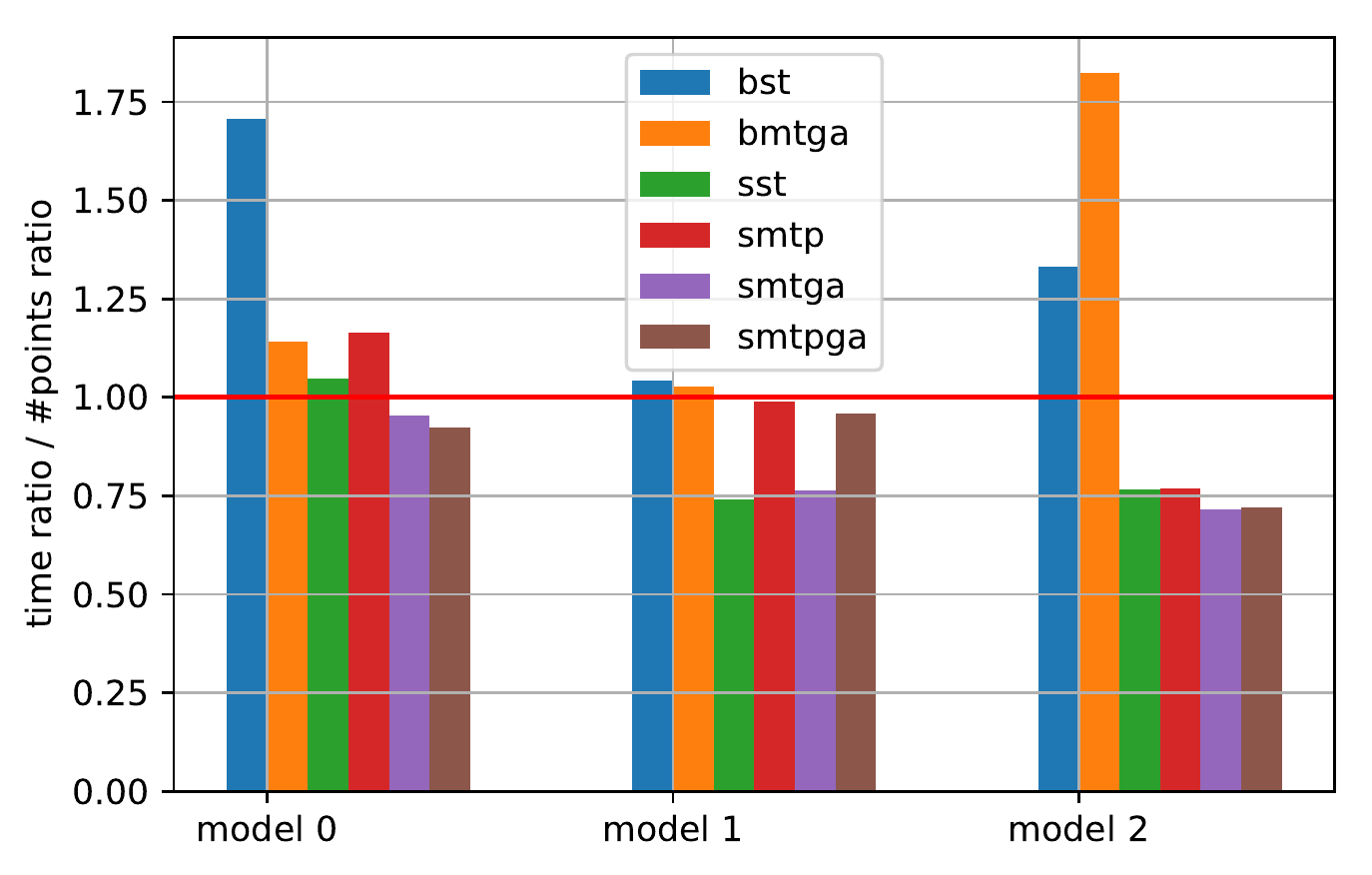}
	\caption{Ratio between high-detail and low-detail point cloud size factor and corresponding timing factors for all models (see Equation \ref{eq:ratio}). The red line indicates linear scaling with a slope of $1$ with respect to point cloud size. Model M$3$ exists only in high-detail.}
	\label{fig:graph3}
\end{figure}

Fig.~\ref{fig:graph3} depicts measurement results for the ratio
\begin{equation} \label{eq:ratio}
\frac{\#\text{points}_{high}}{\#\text{points}_{low}} : \frac{\text{duration}_{high}}{\text{duration}_{low}}, 
\end{equation}
which quantifies the dependency between point cloud size and corresponding computation times.
It indicates that, for larger models (model M$0$ and M$2$), the fastest partitioning approach scales up to $1.9$ times better than the best performing baseline approach with respect to point cloud size.
\section{Conclusion}
\label{sec:conclusion}
In this work, a technique for accelerating an Evolutionary Algorithm for extracting 
a \ac{CSG}-tree from a point cloud was proposed. It is based on a partitioning of the search space obtained 
from computing the maximum cliques of a graph of overlapping primitives, and on merging \ac{CSG}-trees 
extracted for each partition. 
The experimental evaluation indicated a significant speed-up over the baseline approach (the Evolutionary Algorithm) for different modes of parallelization.
\\
One possible direction for future work is 
the implementation of the \ac{GA} for massively parallel computing hardware, combined with the proposed partitioning approach. 
A decreased tree size in the partitioning approach could also be achieved by improving the merge process.
Finally, since the partitioning (and merge) approach described in this work is independent of the technique used for the \ac{CSG}-tree construction, the same approach could potentially be used with the \ac{CSG}-tree conversion approaches in \cite{shapiro1991construction,buchele2004three}. 

	\bibliographystyle{alpha} 
	\bibliography{main} 
\end{document}